\documentclass[aps,prx,twocolumn,showpacs,superscriptaddress,longbibliography]{revtex4-2}

\usepackage{epsfig,graphicx,amsmath,amssymb}
\usepackage{natbib}
\usepackage[nooneline]{subfigure}
\usepackage{physics} 

\usepackage[colorlinks=true, linkcolor=blue, urlcolor=blue, citecolor=blue]{hyperref}

\usepackage{mathtools}

\begin{document}

\title{Fractionally Quantized Recurrence Detection Times in Monitored Quantum Many-Body Systems}

\author{Quancheng Liu}
\email{Corresponding Author: qcliu.ac@gmail.com}
\affiliation{Department of Physics, Institute of Nanotechnology and Advanced Materials, Bar-Ilan University, Ramat-Gan 52900, Israel}
 \author{Sabine Tornow}
\affiliation{Department of Computer Science, Research Institute CODE (Cyber Defence), University of the Bundeswehr Munich, Munich 81739, Germany}
 \author{David A. Kessler}
\affiliation{Department of Physics, Bar-Ilan University, Ramat-Gan 52900, Israel}
\author{Eli Barkai}
\email{Corresponding Author: Eli.Barkai@biu.ac.il}
\affiliation{Department of Physics, Institute of Nanotechnology and Advanced Materials, Bar-Ilan University, Ramat-Gan 52900, Israel}

\begin{abstract}
Recurrence time quantifies the duration required for a physical system to return to its initial state, playing a pivotal role in understanding the predictability of complex systems. In quantum systems with subspace measurements, recurrence times are governed by Anandan-Aharonov phases, yielding fractionally quantized recurrence times. However, the fractional quantization phenomenon in interacting quantum systems remains unexplored. Here, we address this gap by establishing universal lower and upper bounds for recurrence times in interacting many-body spin systems. Notably, we investigate scenarios where these bounds are approached, shedding light on the speed of quantum processes under monitoring.  In specific cases, our findings reveal that the complex many-body system can be effectively mapped onto a dynamical system with a single quasi-particle, leading to integer-quantized recurrence times. Our work demonstrates a valuable link between recurrence times and the number of dark states in the system, thus providing a deeper understanding of the intricate interplay between Hilbert-space fragmentation, ergodicity breaking,  measurements, and interaction effects. Finally, our findings have been implemented on an IBM quantum computer, revealing resonances and fractional quantization in agreement with theoretical predictions. This demonstrates the resilience of non-equilibrium topological fractional quantization to noise and highlights its potential use for benchmarking quantum devices and probing dark states.
\end{abstract}

\maketitle

The concept of first-passage time, referring to the time it takes for a stochastic signal or path to reach a target state, is fundamental in various scientific disciplines~\cite{redner_2001,Metzler_2014,Alan2013,BENICHOU2014225,Gurin2016,Bebon2023}. In quantum physics, this concept is often defined through repeated monitoring of the target state~\cite{Dhar_20152,Gherardini_2016,kulkarni2023detection}, allowing the extraction of valuable information regarding the completion time of quantum processes~\cite{Kempe2005,Dhar2015,Krovi20062,PhysRevB2023}. This measurement-defined first-passage notion is known as the quantum first detection time and has been studied extensively in single-particle dynamics~\cite{Das_2022,Das_20222,Wang2023}. Within this framework, well-known effects include the quantum Zeno limit~\cite{10.1063/1.523304,PhysRevLett.130.103801,PhysRevA.69.032314,Dubey2021},  oscillatory decay of the first detection probability~\cite{Thiel20181}, optimization via quantum resetting~\cite{Kulkarni_2023,Acharya2023,wald2025stochasticresettingdiscretetimequantum}, and resonance phenomena~\cite{Friedman2018,Grnbaum2013}, underscoring how quantum first detection times differ from their classical counterparts~\cite{doi:10.1080/00107151031000110776,Krapivsky2014,doi:10.1073/pnas.2402912121}. In certain cases, the process can display swift detection, a highly sought-after outcome in quantum search~\cite{Kempe2005,Yin2023,PhysRevE.110.034132,PhysRevResearch.7.023069,PhysRevResearch.2.033113}. A more challenging case is the first detection time in many-body systems~\cite{Dhar20063,Agranov2018,Sunghan2023,Dittel2023,walter2023thermodynamic,purkayastha2023interaction,hass2023passage}. Recent work has examined first detection times in fermionic systems~\cite{Dittel2023,walter2023thermodynamic,purkayastha2023interaction}, Ising chains~\cite{Dhar20063}, and classical interacting models~\cite{Agranov2018,Sunghan2023,hass2023passage}. Much of this literature focuses on asymptotic decay of the first detection probability, interpreted through thermodynamic-phase mappings~\cite{walter2023thermodynamic}, measurement-frequency-driven transitions in detection statistics~\cite{Dhar20063,purkayastha2023interaction}, or interference effects in monitored quantum dynamics~\cite{Dittel2023}. By contrast, the interplay between measurements and many-body interactions and the emergence of topological features in first detection times for generic monitored quantum many-body systems remain largely unexplored. 

Another important time scale used to characterize many physical processes is the recurrence time, which measures the time required for a physical system to return to its initial configuration~\cite{Walters1982Ergodic,PhysRevA.93.050101,ROBINETT20041}. In the classical context, the implications of recurrence for thermodynamic irreversibility were debated by Zermelo and Boltzmann, and rigorously treated by Poincar\'e \cite{boltzmann1896vorlesungen,wintner1947analytical}. These studies focused on the deterministic return of conservative systems. Bridging these dynamical foundations with stochastic theory, Kac's lemma links the mean return time to the invariant measure, implying finite expected returns for any set of positive measure~\cite{kac1947notion}. In the quantum domain, recurrence is traditionally viewed as an energy spectrum phenomenon related to state revivals or Loschmidt echoes~\cite{PhysRev.107.337,Krapivsky_2018,stefanak2025recurrence,ROBINETT20041}. For systems with a discrete energy spectrum, the quantum recurrence theorem proves that the state vector will eventually return arbitrarily close to its initial position in Hilbert space, a consequence of the quasi-periodic nature of the underlying unitary evolution.

Combining recurrence with the first detection paradigm, recent theoretical developments have formalized the first detection recurrence time as a first-passage observable: the time to the first successful detection of a return under repeated measurements~\cite{PhysRevA.91.042108,PhysRevA.110.012219,Lahiri2019,Yin2019}. Here the return to the original state plays a special role, since the measurement backaction drives the system along an effectively closed cycle in Hilbert space. This perspective has revealed a connection to the Aharonov–Anandan phase~\cite{Aharonov1987,Bourgain2014}, underscoring the nontrivial interplay between monitoring, quantum interference, and measurement-induced topological effects~\cite{doi:10.1073/pnas.1911620117,Viotti2023geometricphases,e26100869,doi:10.1126/sciadv.adg6810}. More specifically, using the theory of operator-valued Schur functions, Bourgain {\em et al.}~\cite{Bourgain2014} rigorously demonstrated that the recurrence times in a generic Hilbert ${\cal H}$ space with any subspace measurements ${\cal H}_d$ ($\dim {\cal H}_d<\dim {\cal H}$) are fractionally quantized. However, how this elegant but abstract theory manifests in physical systems, especially interacting many-body systems, remains largely unexplored. In this work, we address this gap by revealing how fractional recurrence arises in realistic spin models, establishing universal bounds, linking recurrence times to the structure of dark states, exploring size effects, and experimentally observing fractional quantization using a noisy intermediate-scale quantum computer.

\begin{figure}
    \centering
    \includegraphics[width=0.8\columnwidth]{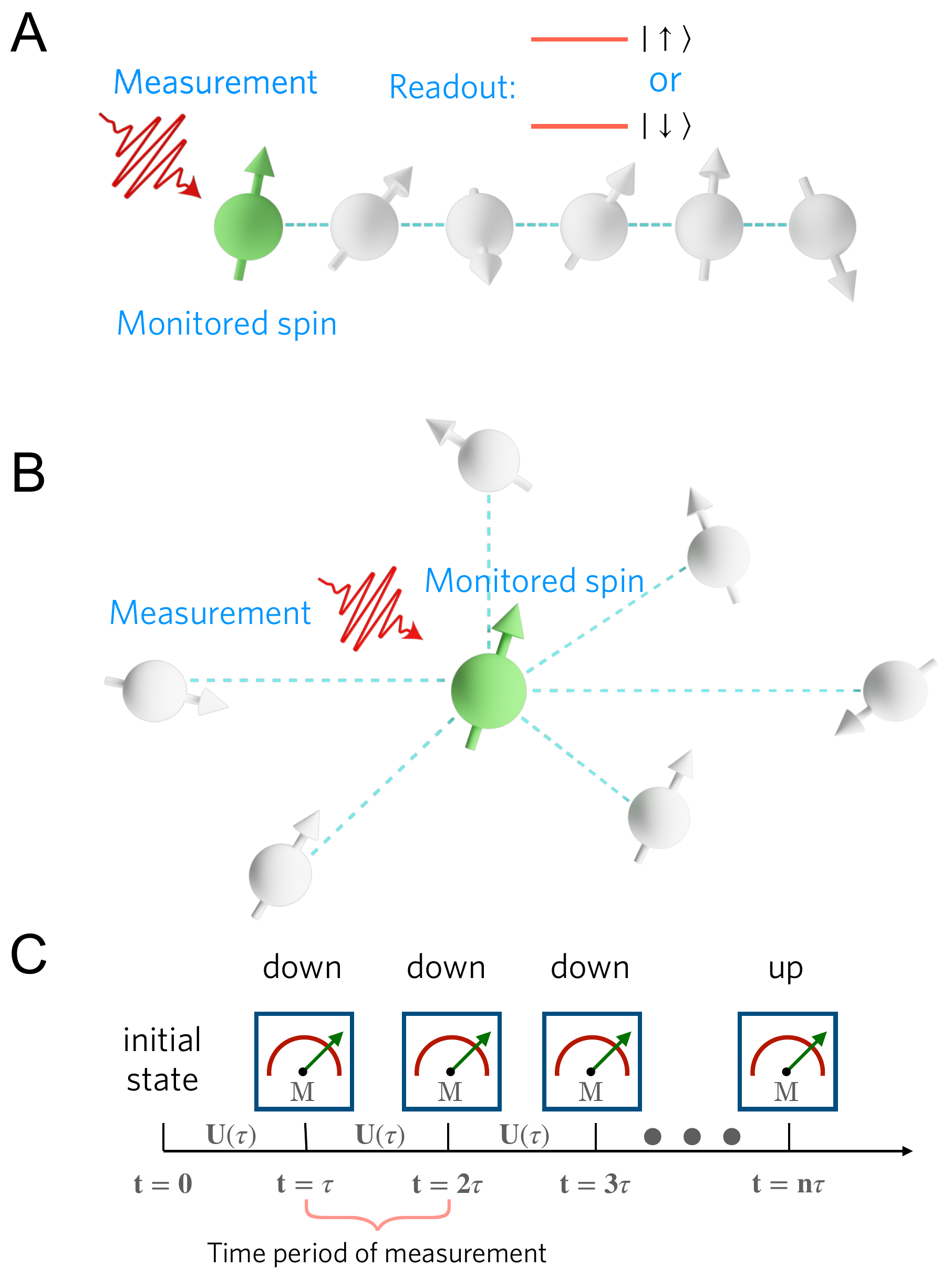}
    \caption{Schematic plots of repeated measurements for two representative models. The Heisenberg spin model (A) and the central spin model (B) with repeated monitoring of a specific spin, which is initially in state ``up". The subspace measurements are performed at discrete times $\tau, 2\tau, 3\tau, \cdots$, where $\tau$ is the time period of measurement. The process ends when we record the monitored spin in the state ``up" for the first time. (C)  The evolution of the systems is a combination of unitary dynamics and the back-action from the measurements.}    
    \label{fig0}
\end{figure}

In our protocol, a single spin or qubit is prepared in the state ``up" and then repeatedly monitored with a time interval $\tau$. The first time to record the spin in the state ``up" is typically random. How the interactions control the statistics of this recurrence time is the basic question.  Studying a wide range of models, we find precise fractional values of the mean recurrence time, $\bar{n}=p/q$, where $q$ is the number of recurrence initial states and $p$ counts the bright states in the monitored recurrence dynamics, as clarified below. We explore how the fractional recurrence time behaves in the thermodynamic limit and obtain universal upper and lower bounds, also valid far from this limit. One of our goals is to relate the fractional recurrence time and the number of dark states in the system. Such a relationship will demonstrate how recurrence times are connected to the fragmentation of Hilbert space under repeated measurements and hence to ergodicity breaking. We also address the control of the fractional ratio $p/q$ by finite measurement time effects, which allows for characterizing spin environments in the laboratory. In addition, we uncover resonances in the recurrence time where, for special values of $\tau$, the ratio $p/q$ drops significantly.  Finally, we validate our theory on an IBM quantum computer, observing the fractional quantization and its robustness to noise.  We also introduce a novel quantum speed-up method that utilizes an entanglement-assisted protocol to measure the mean recurrence time without performing an exponentially large number of averages. This establishes recurrence statistics as a viable tool for benchmarking mid-circuit measurements in noisy realistic hardware.

\begin{figure*}
    \centering
    \includegraphics[width=1.5\columnwidth, trim=0cm 0cm 0cm 0cm, clip]{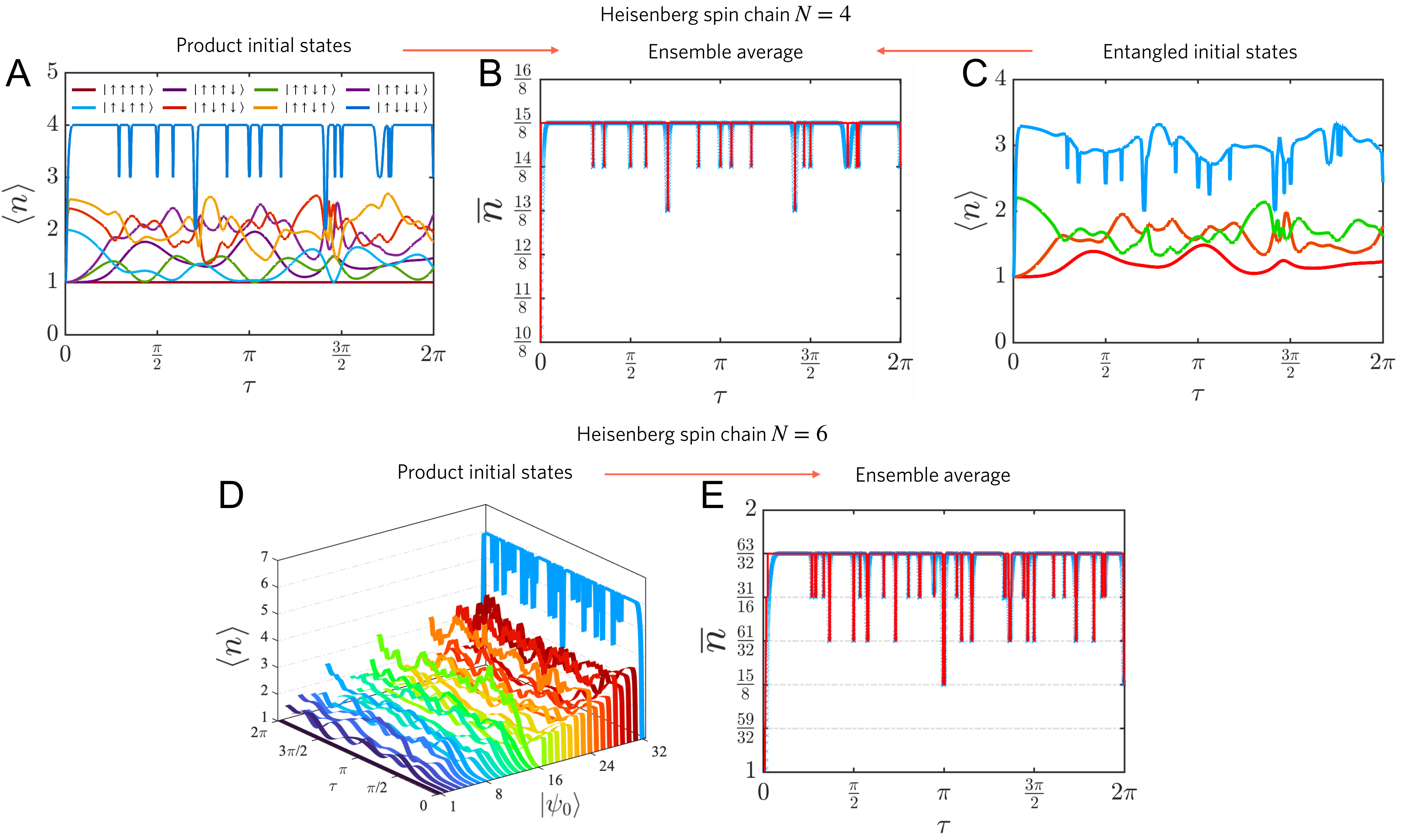}
    \caption{Mean recurrence time for the Heisenberg spin chain. (A) $\langle n \rangle$ for product initial states with four spins. (B) The corresponding ensemble mean $\overline{n}$ versus time interval $\tau$. The mean is $15/8$ except for special values of $\tau$, where it is $14/8$, $13/8$ or 1 when $\tau\rightarrow 0$. The blue crosses are numerical simulations [averaging of all states in (A)], which perfectly match the theoretical red line given by Eq. (\ref{eq09}). (C) $\langle n \rangle$ for entangled initial states with four spins. While plots (A) and (C) differ significantly, averaging over all states yields fractional recurrence times that are independent of the initial basis choice, as shown in plot (B). (D) Same as (A) for six spins, where now we have 32 product states as initial states. (E) The six-spin theoretical (red line) and numerical (blue crosses) $\overline{n}$ perfectly match and exhibit fractional quantization. For most $\tau$, $\overline{n}=63/32$. In the numerical $\langle n \rangle$ and $\overline{n}$, the resonances are of finite width. This is because the number of measurements in the numerical simulations is finite. If we increase this number, the width will narrow and in the limit, the resonances become point-wise discontinuities.    The figure clearly demonstrates the rich physical behavior that emerges from specific initial conditions. Remarkably, this complexity reduces to a simple and elegant structure when considering the mean $\bar{n}$, which exhibits fractional quantization, a striking feature that not only reflects the underlying symmetry but also reveals the number of dark states in the system, as shown below.}    
    \label{fig1}
\end{figure*}

\section{Preliminaries}

\subsection{Measurement protocol}
 We explore the dynamics of $N$ interacting spins governed by a time-independent Hamiltonian $H$, e.g., the Heisenberg spin model, central spin models, Ising model, and the Heisenberg spin model with Dzyaloshinskii-Moriya (DM) interactions \cite{PhysRevB.52.10239}. However, the general formalism we develop here is valid for any interacting spin model. To illustrate, consider the Hamiltonian $H$ for the Heisenberg spin model, given by:
\begin{equation}
H = \sum_{\langle i, i+1\rangle } \sum_{\alpha =x,y,z} S_{i}^{\alpha} S_{i+1}^{\alpha} ,
\label{eqN101}
\end{equation}
where $S_i^\alpha$ denotes the spin-$\frac{1}{2}$ operators. We now add the component of repeated measurements, which in generality may yield many novel and sought-after outcomes \cite{PhysRevX.9.031009,PRXQuantum.2.010352,PhysRevLett.122.070603,PhysRevE.83.041114,PhysRevE.98.032108,PhysRevResearch.2.013095,PhysRevLett.132.010402,PhysRevB.111.224305,Bhattacharya2024}. We consider the measurement of a single spin and the other non-monitored spins comprise the ``bath". Below we will consider finite $N$ and large $N$ limits, so the ``bath" does not necessarily have to be vast. The measurements are performed at times $\tau$, $2 \tau$, $3 \tau$, etc, and in the time interval between the measurements, the system evolves unitarily with $U(\tau) = \exp(-i H \tau)$ and $\hbar=1$, see Fig. \ref{fig0} for schematics of the measurement process. Each measurement employs a projection in the $Z$ direction, determining the state of the observed spin as either ``up" $\uparrow$ or ``down" $\downarrow$. The states of the unobserved spins remain unknown; hence, we obtain partial information about the whole system, making the concept of subspace measurements suitable. It is important to note that due to the entanglement between the spins, the effects of measurements are spread to the whole system. As a result, the many-body system propagates under the combined effects of the unitary dynamics and the back-action of measurements.

We study the first detection recurrence time of the monitored spin. Specifically, the monitored spin is initialized in $\ket{\uparrow} $ at $t=0$, we measure whether it is in state $\ket{\uparrow}$ or $\ket{\downarrow}$, and stop the process when the first ``up" is recorded. For example, we may record a sequence $\{ \mbox{down}, \mbox{down}, \mbox{up}\}$ at times $\{ \tau, 2\tau, 3\tau \}$  so that $3 \tau$ is the first detection recurrence time. We denote $n$ the number of measurements, and  $F_n$ the probability of detecting the spin in the ``up" state for the first time in the $n$-th attempt. Notably, this is a local/subspace notion of recurrence: at $n\tau$ the monitored spin has returned to its initial state, while the remaining spins may be different.  We focus on the mean $\langle n \rangle$, which determines the mean recurrence time till the first detection $\langle n\rangle \tau$. As our quantum simulation below demonstrates, such single-spin measurements can be implemented on quantum computers~\cite{Klaus2023,Cech2023,kamakari2024experimental,PhysRevResearch.6.013311,Koh2023}, where qubits replace spins.

This measurement-based recurrence differs from the classical Poincaré recurrence time~\cite{wintner1947analytical,SINGH2022133418}, and also from its quantum counterpart for closed systems with discrete spectra~\cite{PhysRev.107.337}, where purely unitary evolution guarantees returns arbitrarily close to the initial state. It is also distinct from quantum recurrences defined via revivals and Loschmidt echoes~\cite{GORIN200633}, where one characterizes return under purely unitary evolution through conditions such as $|\langle \psi(t)|\psi(0) \rangle |^2=1$. In contrast, the recurrence time here is a measurement-defined first-passage quantity and is generally stochastic. Nevertheless, the intrinsic frequencies of $H$ still shape the statistics of first detection recurrence time, leading to the resonant features discussed below.

\subsection{Example 1: Heisenberg spin chains} The measurement protocol we use leaves a plethora of choices for the initial state of the bath spins. Starting with the Heisenberg model, we plot the mean first detection recurrence time $\langle n \rangle$ versus the time interval between measurements $\tau$ for a four-spin chain system (\textit{Materials and Methods}). The measurements are performed on the leftmost spin. As shown in Fig. \ref{fig1}, we consider $2^{N-1}$ initial product states, and since $N=4$, we have in total 8 cases. We witness three types of initial conditions. A trivial case is when all bath spins are initially ``up", and then $\langle n \rangle =1$. This initial condition is a stationary state of $H$, and hence the state ``up" of the monitored spin is recorded with probability one in the first measurement. For other initial states, the average is a rather intricate function of the time period between measurements, though in all cases the mean does not exceed $3$, hence the mean detection time is not exceedingly long. Finally, for initial state $\ket{\psi_0}= \ket{\uparrow,\downarrow,\downarrow,\downarrow}$, we attain the value $\langle n \rangle=4$ except for resonances, found for special choices of $\tau$. We see that in this case the mean $\langle n \rangle$ is integer-quantized, either $1$ for the Zeno limit $\tau \to 0$~\cite{PhysRevLett.86.2699}, or $2$, or $3$, while typically it is $4$. Such an initial state exhibits what we call the integer mean recurrence time. The resonances in $\langle n \rangle$ are of particular interest, as they reflect underlying periodicities in the quantum system and are also observed, in a regularized form, in the quantum computer simulation results analyzed below.

While the sensitivity of the results on the specific initial conditions is clearly seen, when we average uniformly over all the initial conditions, a remarkably smooth result is found. In Fig. \ref{fig1}(B), the ensemble mean denoted $\overline{n}$,  for most $\tau$, is equal to $15/8$. The intricate behaviors of $\langle n \rangle$ are now replaced by a topologically protected behavior encapsulated in the ensemble mean $\overline{n}$, which is robust as we change the system parameter $\tau$. We also see resonances after averaging; for example, as we vary $\tau$, $p$ jumps from $15$ to $14$ and back at the same $\tau$ values as for the special initial state $\ket{\uparrow, \downarrow,\downarrow, \downarrow}$. We also show in Fig. \ref{fig1}(D,E) that the fractional mean for $\bar{n}$, along with the integer mean $\langle n \rangle$ for specific initial conditions, persists for larger systems. We will soon study the basics of fractional quantization of $ \overline{n}$ in general and later return to the integer mean recurrence time, which, as mentioned, is found for specific initial conditions.


 \subsection{Example 2: Entangled initial states} In Fig. \ref{fig1}(A,D), we considered initial states that are product states. However, fractional quantization can be found when considering other choices. In Fig. \ref{fig1}(C), we use a set of entangled initial states that are orthogonal and span the bath space. The explicit expression for the entangled initial states is given in the SI Appendix, section 1. Now, unlike the behavior in Fig.  \ref{fig1}(A,D), resonances are no longer limited to a unique initial state. Yet, the fractional quantization found after averaging remains unchanged, see Fig. \ref{fig1}(B). Below, in our examples, we will employ product initial states, though our main conclusions remain unchanged for other basis.

\subsection{Example 3: Dzyaloshinskii-Moriya interactions}
The quantization behavior of $\bar{n}$ extends beyond normal Heisenberg spin chains and persists in more realistic cases incorporating Dzyaloshinskii-Moriya (DM) interactions~\cite{CAMLEY2023100605}. These interactions introduce spin-orbit coupling effects that can significantly alter the system's symmetry properties. The specific form of the DM Hamiltonian and detailed numerical results are provided in SI Appendix, section 2 and Fig. S1. Varying the DM vector $\vec{D}_{\mathrm{DM}}$ yields both fractional plateaus and integer values (e.g., $\bar{n}=2$), with the latter occurring in strongly asymmetric regimes that suppress the dark state manifold as we demonstrate below. Notably, in all cases considered, a resonance in $\bar{n}$ emerges in the Zeno limit $\tau \to 0$, suggesting that this feature is generic.

\subsection{Basic questions raised by the examples} How can the values of $p$ and $q$ be determined from fundamental spin models? What are the physical meanings of $p$ and $q$? Then, what are the upper bounds and lower bounds for recurrence times? What are the thermodynamic limits of the problem? When do we have fractional quantization, and when do we see integer values? What is the physical process responsible for resonances? Will fractional quantization and resonances be detectable using remote simulations on a noisy quantum computer?  Finally, can the averaging process required to determine $\bar{n}$ be avoided by using a single configuration? This is a crucial question since detecting $p$ and $q$ in large systems requires an exponentially large number of initial conditions. However, it will be shown that with resources scaling only linearly with system size, and by using ancillas, this goal can be achieved.

\section{Formalism}
In Eqs. (\ref{eq01}-\ref{eq05}) we recap the basic mathematical formalism underlying our analysis~\cite{Bourgain2014}. While the mathematical formalism was rigorously established by Bourgain \textit{et al.}~\cite{Bourgain2014} using the theory of operator-valued Schur functions, here we identify the physical mechanism governing this phenomenon in interacting many-body systems. By evaluating the generating function in terms of the survival operator, we reveal that the fractional value is explicitly determined by the counting of dark states---eigenstates that remain trapped in the undetected subspace. This connection transforms the topological invariant into a physical probe for Hilbert space fragmentation and ergodicity breaking, enabling us to derive universal bounds on recurrence times and predict measurement-induced resonances.

The measurement of the spin is described by the projection  $D^{\uparrow} = \sum_{b} \ket{\uparrow, b} \bra{\uparrow, b}$ and its complement $D_{\downarrow} = 1 - D^{\uparrow}$, where $b$ stands for bath. The probability of detecting the spin in the state ``up" for the first time in the $n$-th measurement is
\begin{equation}
F_n = || M_n \ket{\psi_0}||^2, \mbox{with}\ \  M_n = D^{\uparrow} U (D_{\downarrow} U )^{n-1}.
\label{eq01}
\end{equation}
The operator $M_n$ represents the mix of measurements and unitaries $U=\exp(-i H \tau)$, and indicates that $n-1$ measurements failed to detect the spin in the target state, followed by a final success. Analytically, $\langle n \rangle$ is found using a generating function technique. Let $\hat{M}(e^{i \theta})= \sum_{n=1} ^\infty M_n \exp( i n \theta)$, then summing the resulting geometric series
\begin{equation}
\hat{M} (e^{ i \theta}) = e^{i \theta} D^{\uparrow} U \left( 1 - e^{ i \theta} D_{\downarrow} U\right)^{-1}.
\label{eq02}
\end{equation}
The operators $M_n$ can then be retrieved via Cauchy's integral formula:
\begin{equation}
	M_n =\frac{1}{2\pi i }\int_0^{2\pi} \hat{M} (e^{i \theta})e^{-i n\theta} d \theta.
	\label{neq1}
\end{equation}
Using the identity $ \int_0 ^{ 2 \pi} \exp[ i (m-n) \theta] {\rm d} \theta= 2\pi \delta_{mn}$, we find
\begin{equation}
\langle n \rangle 
=\frac{1}{2\pi } \int_0^{2\pi} \bra{\psi_0}  \sum_{k=1}^\infty M_k^{\dagger} e^{i\theta k} \left(-i \frac{\partial}{\partial \theta} \right)  \sum_{l=1}^\infty M_l e^{-i\theta l} \ket{\psi_0} {\rm d}\theta.
\end{equation}
Employing the definition of the generating function $\hat{M}(e^{i \theta})$, the mean first detection recurrence time for initial state $|\psi_0\rangle$ is given by~\cite{Bourgain2014}:
\begin{equation}
\langle n \rangle = { 1 \over 2 \pi i} \int_0 ^{ 2 \pi}
\bra{\psi_0} \left[ \hat{M} \left(e^{ i \theta} \right)\right]^{\dagger} \partial_\theta \hat{M}\left(e^{ i \theta} \right) \ket{\psi_0} {\rm d} \theta. 
\label{eq03}
\end{equation}
Eq. (\ref{eq03}) gives the mean recurrence time for a specific initial state $\ket{\psi_0}$. For example, we plot $\langle n \rangle$ versus $\tau$ for different $\ket{\psi_0}$ in Fig. \ref{fig1}(A) and Fig. \ref{fig1}(B) for the Heisenberg model, and as mentioned $\langle n \rangle$ can exhibit a range of behaviors from chaotic-like, to a constant with pointwise resonances, or the trivial behavior $\langle n \rangle=1$.

Next, we calculate the ensemble mean recurrence time $  \overline{n} $, which involves averaging over all possible initial recurrence states. By definition, the ensemble average recurrence time is expressed as:
\begin{equation}
	 \overline{n}  = \frac{\sum_{b} \langle n\rangle_{|\psi_0\rangle=\mid\uparrow,b\rangle }}{\mbox{number of initial states } |\psi_0\rangle=\mid\uparrow, b \rangle },
	 \label{eqn2}
\end{equation}
where $\langle n\rangle_{\mid \uparrow, b \rangle }$ is the mean recurrence time for the recurrence initial state $|\psi_0\rangle = \mid\uparrow, b\rangle$. While we often employ product states for convenience, the linearity of the formulation ensures that the result is basis-independent. Consequently, the fractionally quantized value is basis independent, as demonstrated in Fig. \ref{fig1}(C) using entangled initial states and by comparison with product initial states in Fig. \ref{fig1}(A). Using Eqs. (\ref{eq03}) and (\ref{eqn2}), the ensemble mean recurrence time, averaged over the bath states, is 
\begin{equation}
 \overline{n}  = { 1 \over 2^{N-1}}  \sum_{b} \int_0 ^{ 2 \pi} \bra{ \uparrow, b} 
\left[ \hat{M} \left( e^{ i \theta} \right) \right]^{\dagger} \partial_\theta \hat{M} 
\left(e^{ i \theta} \right) \ket{\uparrow, b} {{\rm d} \theta \over 2 \pi i},
\label{eq04}
\end{equation}
The value of $q$ is thus the denominator $2^{N-1}$, hence for $N=4$ we have in Fig. \ref{fig1}(B) $q=8$ while in Fig. \ref{fig1}(E) $N=6$ and $q=32$. Eq. (\ref{eq04}) includes a trace over the bath states, since we sum matrix elements over all the $2^{N-1}$ bath states.  We can rewrite this as:
\begin{equation}
	 \overline{n} = { 1 \over 2^{N-1}}   \int_0 ^{ 2 \pi} \mbox{Tr} \left( \left[ \hat{M} \left( e^{ i \theta} \right) \right]^{\dagger} \partial_\theta \hat{M} 
\left(e^{ i \theta} \right) \right) {{\rm d} \theta \over 2 \pi i}.
\label{eq0000}
\end{equation}
From Eq.~(\ref{eq04}), the summation extends only over the states \(\ket{\uparrow, b}\). Here, however, one traces over all possible states. This works because of \(D^{\uparrow}\) in the generating operator, as specified in Eq.~(\ref{eq02}). In evaluating the trace, the remaining states \(\ket{\downarrow, b}\) contribute nothing and thus yield zero. Consequently, it is valid to interpret this partial summation as the \(\mathrm{Tr}\) of the generating operator.

We now define the eigenvalues and eigenvectors of generating function $\hat{M} \left( e^{ i \theta} \right)$ as: $\hat{M}(e^{i\theta}) |m_k(\theta)\rangle = m_k(\theta) |m_k(\theta)\rangle$. Expanding the $\left[ \hat{M} \left( e^{ i \theta} \right) \right]^{\dagger} \partial_\theta \hat{M} 
\left(e^{ i \theta} \right)$ in this eigenspace, the trace becomes
\begin{equation}
\begin{aligned}
	\mbox{Tr} \left( \left[ \hat{M} \left( e^{ i \theta} \right) \right]^{\dagger} \partial_\theta \hat{M} 
\left(e^{ i \theta} \right) \right) &= \sum_k m_k^*(\theta) \partial_\theta m_k(\theta) \\
&=  \partial_{\theta} {\rm ln} \left[  {\rm det} \left[
\hat{M} \left( e^{ i \theta} \right) \right] \right],
\end{aligned}
\end{equation}
where ${\rm det} \left[ \hat{M} \left( e^{ i \theta} \right) \right]$ is the determinant of the generating function and $*$ denotes the complex conjugate. The determinant of $\hat{M}(e^{i \theta})$ is the product of its eigenvalues, some of which are zero. An epsilon regularization is implied: each zero eigenvalue is replaced by a small regulator $\epsilon$, and the limit $\epsilon \to 0$ is taken with $\partial_\theta \epsilon / \epsilon = 0$. In this way, the zero eigenvalues do not contribute to the logarithmic derivative and are effectively excluded.  Using Eq. (\ref{eq0000}), the ensemble mean recurrence time reads~\cite{Bourgain2014}
\begin{equation}
 \overline{n}  = { 1 \over 2^{N-1} } \int_0 ^{ 2 \pi} { \partial_{\theta} {\rm det} \left[
\hat{M} \left( e^{ i \theta} \right) \right] \over {\rm det} \left[ \hat{M} \left( e^{ i \theta} \right) \right]}
{{\rm d} \theta \over 2 \pi i} . 
\label{eq05}
\end{equation}
%
To find $p$, we use ${\rm det} (ab)= {\rm det} (a){\rm det} (b)$ and ${\rm det} (a^{-1})= 1/{\rm det} (a)$ and find
\begin{equation}
{\rm det} \left[ \hat{M}\left(e^{ i \theta} \right) \right] = {\rm det} [D^{\uparrow} U]
{1 \over {\rm det} [ \exp( - i \theta) - D_{\downarrow} U]}.
\label{eq06}
\end{equation}
The term ${\rm det} [D^{\uparrow} U]$ is $\theta$ independent and using the $\theta$ derivative in  Eq. (\ref{eq05}) this term is clearly not contributing to $\overline{n}$.  We now define the survival operator $S= D_{\downarrow} U$ whose eigenvalues are denoted $\xi_j$, so ${\rm det} (S - \xi)=0$ and they satisfy $|\xi_j|\le 1$. Using Eqs. (\ref{eq05},\ref{eq06}), we can then write
\begin{equation}
\overline{n} = 
{1 \over 2 q\pi i} \int_{0} ^{ 2 \pi}
{ \partial_\theta \ln  \left[ \Pi_{j=1} ^{2^N} \left( e^{-i \theta} -\xi_j\right)\right]^{-1} 
{\rm d } \theta}
\label{eq07}
\end{equation}
and $q=2^{N-1}$ as mentioned. Using 
\begin{equation}
\int_0 ^{ 2 \pi} { e^{ - i \theta} \over e^{-i\theta} - \xi_j} {\rm d} \theta= \left\{
\begin{array}{c c} 
2 \pi & |\xi_j|<1 \\
0 & |\xi_j|=1
\end{array}
\right.
\label{eq08}
\end{equation}
we find 
\begin{equation}
\overline{n}= {  \mbox{number of} \  \xi_j \ \mbox{in the unit circle} \over 2^{N-1} }.
\label{eq09}
\end{equation}
Hence, $p$ is the number of eigenvalues of the survival operator $S$ whose absolute values are less than unity. The resonances, examples of which are presented in Fig. \ref{fig1}(B,E), are found whenever an eigenvalue of $S$ approaches the unit circle as we vary the $\tau$ (see SI Appendix, Fig. S2 for details).


The number of $\xi$s equal to zero is at least $2^{N-1}$, and hence $1 \le \overline{n}$, which is expected. Since we have in total $2^N$ eigenvalues of $S$ for any spin-half model, we find the universal bound
\begin{equation}
 1 \le \overline{n} \le 2.
\label{eqTB}
\end{equation}
Such a result is remarkable as the upper bound on the mean number of measurements is independent of the details of $H$. In some sense, the formula is in the spirit of a speed limit~\cite{Mandelstam1991,MARGOLUS1998188,Sun2019,Vu2023}, though these discuss how two states evolve, while we consider detecting a target state with subspace measurements. The lower bound is reached as an equality for the Ising model on a chain,  while the upper bound is found for the Ising model with measurements either in the $X$ or $Y$ direction for typical values of $\tau$. For strongly disordered systems with no underlying symmetries, we expect \( \bar{n} = 2 \). Indeed, this was demonstrated in the strongly anisotropic DM model.  This follows from our observation below that \( p \) is determined by the number of dark states in the system.

\begin{figure}
    \centering
    \includegraphics[width=1\columnwidth, trim=0cm 0cm 0cm 0cm, clip]{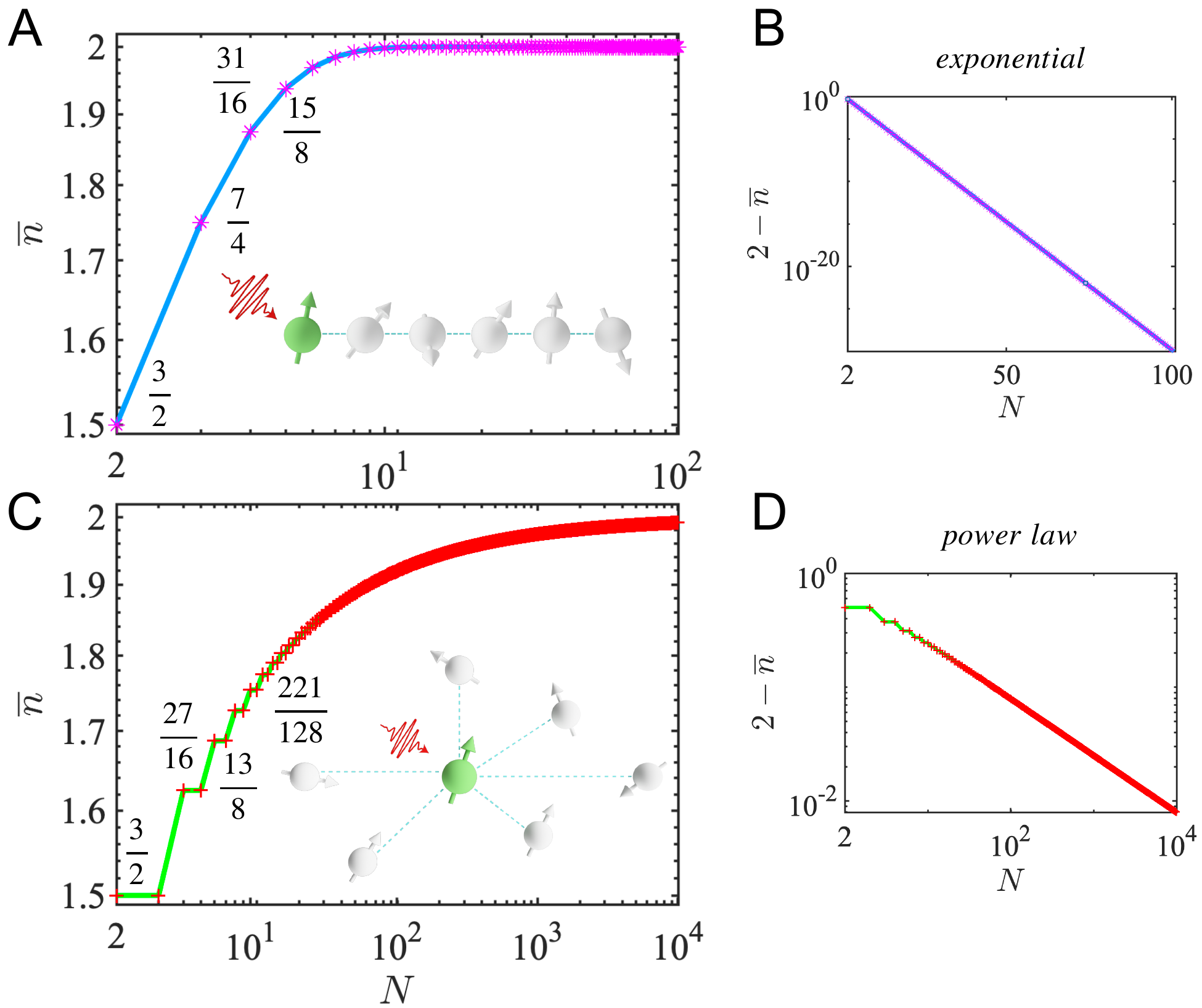}
    \caption{Mean recurrence time as a function of system size $N$. (A,B) Heisenberg spin chain: (A) $\bar{n}$; (B) the exponential fall of $2-\bar{n}$  in the thermodynamic limit. (C,D) XX central spin model: (C) the staircase growth of the fractionally quantized $\bar{n}$; (D) the power-law fall of $2-\bar{n}$ in the thermodynamic limit.}    
    \label{fign2}
\end{figure}

\section{Dark States Give Fractional Quantization}

 Eq. (\ref{eq09}) is very useful in the evaluation of $\overline{n}$, yet it does not address the main physical aspect of the problem. We claim that $\overline{n}$ provides information on the number of dark states in the system. Dark states $\ket{D}$ are initial conditions that are never detected~\cite{Thiel2020,Plenio1998}. Roughly speaking, under repeated measurements, the Hilbert space is divided into dark and bright subspaces, and it exhibits fragmentation related to ergodicity breaking. We now explain this novel connection between recurrence times and dark states.

Define the eigenvalue problem $S\ket{D} = \xi \ket{D}$. We consider the eigenvalues that are {\em on} the unit circle, so $\xi= \exp(-i \gamma)$. It follows that $S^{n-1} \ket{D} = \exp[ - i \gamma(n-1)] \ket{D}$. Using Eq. (\ref{eq01}) with $|\psi_0\rangle=|D\rangle$ we have $F_n= \lVert D^{\uparrow} U \exp[ - i \gamma(n-1) ] \ket{D}\rVert^2$, which is clearly independent of $n$. From the basics of probability, this is possible only if $F_n=0$; hence the state $\ket{D}$ is never detected, i.e., it is dark. It then follows, using $\lVert D^{\uparrow} U \ket{D}\rVert^2=0$, that $U\ket{D} = S \ket{D}$, and so $U\ket{D} = \exp( - i \gamma \tau) \ket{D}$. The eigenvalues of $U$ are well-known and given by the exponentials of $-i\tau$ times the energies. We thus conclude that the eigenstates of the survival operator, corresponding to eigenvalues on the unit circle $|\xi|=1$, are stationary energy states of the Hamiltonian $H$ with the measured spin in the state ``down". As mentioned, if the initial condition is chosen to be one of these states, the spin is never found in the target state ``up". Thus, the recurrence problem, which deals with bright initial conditions, i.e., states that are detected with probability one, yields information on the number of dark states in the system and the fragmentation of the Hilbert space. Using Eq.~(\ref{eq09}) we find
\begin{equation}
 \overline{n} = 2 - { \mbox{number of dark states} \over 2^{N-1}}.
 \label{eq11}
\end{equation}
We see from the examples of the Heisenberg and DM models that the latter can have no dark states, while the former does. For strongly anisotropic systems like the DM model, the symmetry breaking destroys dark states, meaning the Hilbert space does not decompose into dark and bright submanifolds, a behavior that is reminiscent of classical Markovian systems, which are ergodic and have no dark subspace.
 
From here, we can also infer the resonances, which are found for those special choices of $\tau$ where a dark state is created. Consider two non-degenerate energy eigenstates, $\ket{E_1}$ and $\ket{E_2}$, for which the condition holds:
\begin{equation}
	D^{\uparrow} (\alpha_1 |E_1\rangle + \alpha_2 |E_2\rangle) = 0.
	\label{eqN18}
\end{equation}
 Then, for special $\tau$s matching $\exp(- i E_1 \tau)=\exp( - i E_2 \tau)$, the number of dark states can increase by unity, and hence we will experience a resonance in $\overline{n}$, as shown already in Fig. \ref{fig1}. Further examples of dark states that lead to resonances in $\overline{n}$ are discussed in SI Appendix, section 4, and Fig. S3 for details.  Hence, a necessary condition for resonances is the phase-matching condition, which resembles quasi-periodicity of two energy states.  In the quantum Zeno limit, the unitary approaches the identity, $U(\tau)\rightarrow I$. The survival operator then reduces to $S=D_{\downarrow}$, so every state with the monitored spin in the state ``down” is dark, yielding $2^{N-1}$ dark states. Accordingly, the mean recurrence time reaches its lower bound, $\bar{n}=1$, consistent with the resonances at $\tau \rightarrow 0$ seen in Figs. \ref{fig1} and \ref{fig:XYN20}. Away from the Zeno limit, including typical or large values of $\tau$, the dark state manifold is unchanged except at isolated resonant points. Thus, increasing $\tau$ does not generically enlarge or reduce the dark space; instead, it produces sharp changes only when exact phase matching occurs. In practice, large $\tau$ can introduce additional effects in quantum computer implementations, since longer intervals typically expose the system to higher noise levels.

\section{Scaling of Recurrence Time with System Size}

In Fig. \ref{fig1}, the recurrence time is $15/8$ for $N=4$ and $63/32$ for $N=6$, excluding resonances. So how does $\overline{n}$ scale with the system size? Or more fundamentally, how does $\overline{n}$ behave in the thermodynamic limit? To address this, we investigate the behavior of dark states in different models. In our discussion below, we exclude the resonant $\tau$s; namely, we focus on the typical value of the recurrence time. One exception is the Zeno limit $\tau \rightarrow 0$, where we expect $\bar{n} =1$, which should be considered with care. Our goal is to show that the approach to the thermodynamic limit is non-universal.

For the Heisenberg chain model, our analysis reveals a single dark state, characterized by all spins being ``down". This state, with an eigenenergy of $(N-1)/4$, leads to $S\ket{D} = \exp[- i (N-1) \tau /4]\ket{D}$. Using Eq. (\ref{eq11}),
\begin{equation}
	\overline{n} = 2- \frac{1}{2^{N-1}},
	\label{eqN111}
\end{equation}
as depicted in  Fig. \ref{fign2}(A). Here, then, $\overline{n}$ exponentially approaches the upper bound, see illustration in Fig. \ref{fign2}(B).

\begin{figure*}
    \centering
    \includegraphics[width=1.6\columnwidth, trim=0cm 0cm 0cm 0cm, clip]{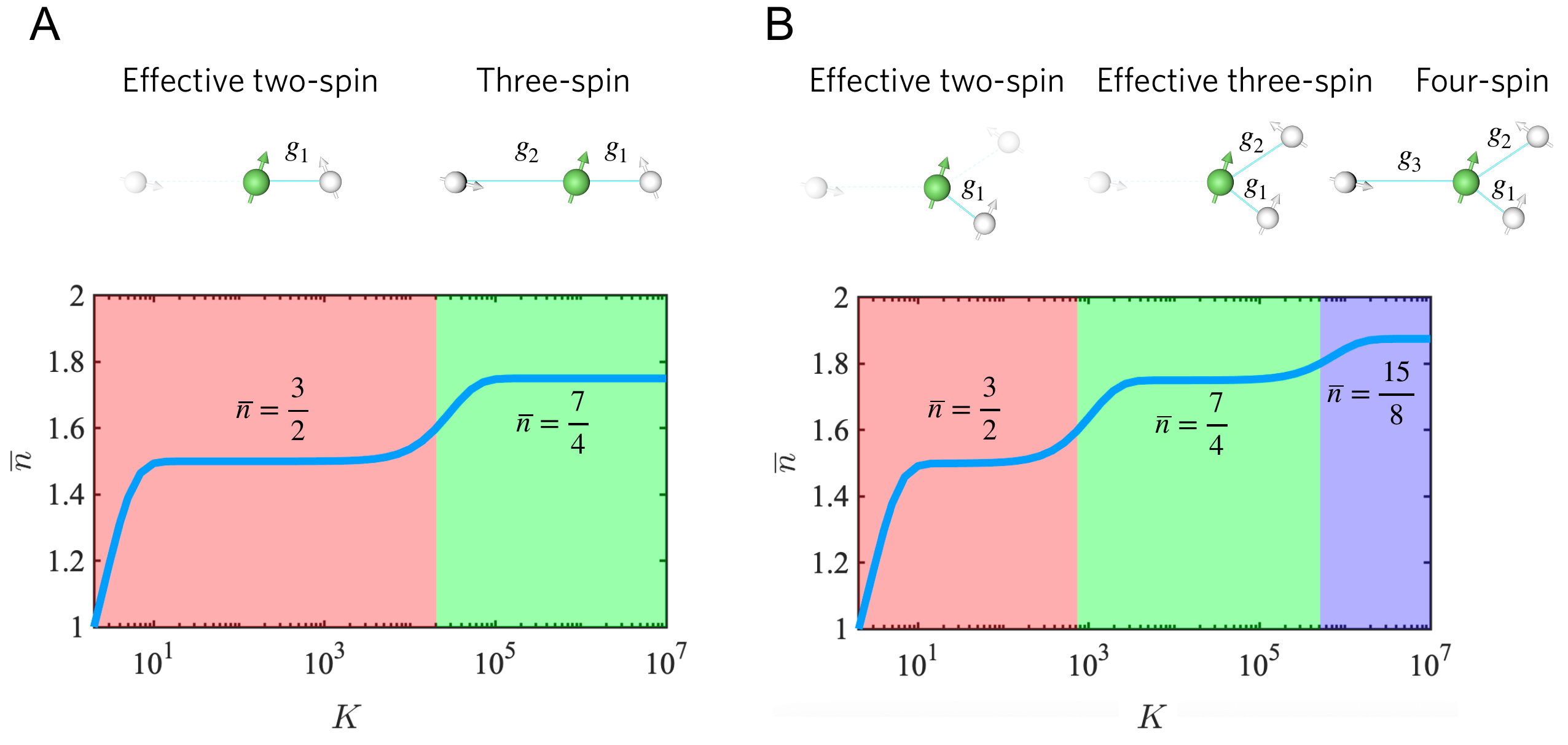}
    \caption{$\overline{n}$ versus the number of measurements reveals fractional staircases. (A) The XXX central three-spin model with $g_1 =1$ and $g_2 = 0.01$. Two substantial plateaus are visible within $\overline{n}$, which corresponds to an effective two-spin system for small $K$, followed by a transition to the three-spin model for large $K$. (B) The four-spin model. The weakly coupled spins are gradually contributing to $\overline{n}$ as $K$ increases. The figure shows that finite time fractional quantization is valuable for probing local environments.}    
    \label{fign3}
\end{figure*}

A distinct behavior is found for the XX central spin model, where $H= w_0 S_0^z +w \sum_{i=1}^{N-1} S_i^z + \sum_{i=1}^{N-1} g_i (S_0^+S_i^-+S_0^-S_i^+)$~\cite{Villazon2020}, describing a central spin in a magnetic field $w_0$, interacting with bath spins subject to a magnetic field $w$. Here $S_0^{\pm}$ and  $S_i^{\pm}$ are the spin ladder operators; the index $0$ labels the central spin and $i$ labels a bath spin. The terms $S_0^+S_i^-$ and $S_0^-S_i^+$ describe spin-exchange processes between the central spin and bath spin $i$. Here, the condition for the dark states is $\sum_{i=1}^{N-1} g_i S_i^- \ket{D} = 0$, and all the dark states are degenerate with eigenenergy $E=-\omega_0/2$. As shown in SI Appendix, section 5, the number of dark states grows with the size of the system. Using Eq. (\ref{eq11}), we have  
\begin{equation}
	\overline{n}  = 2 -  \frac{\Gamma (N)}{\Gamma \left( \lfloor\frac{N+2}{2}  \rfloor \right) \Gamma \left( \lfloor\frac{N+1}{2}  \rfloor \right)2^{N-1}}.
	\label{eq12} 
\end{equation}
Here $\Gamma(N)=(N-1)!$ and $\lfloor x\rfloor$ denotes the floor function. As depicted in Fig. \ref{fign2}(C), the mean recurrence time versus $N$ is fractionally quantized and exhibits a unique staircase growth structure. Remarkably, Eq. (\ref{eq12}) is valid for any value of the coupling $g_i$. We next examine the rate at which $\overline{n}$ approaches 2. Using Stirling's formula, the Gamma function expansion for odd and even $N$ yields
\begin{equation}
 2- \overline{n} \sim \sqrt{ 2/(\pi N) }.
	\label{eqN222}
\end{equation}
 This indicates a power-law approach towards the upper bound, as shown in Fig. \ref{fign2}(D), contrasting with the exponential behavior observed in Fig. \ref{fign2}(B). Comparing Eqs. (\ref{eqN111}) and (\ref{eqN222}), for the Heisenberg and central spin models respectively, the vastly different behaviors are due to the large number of dark states found for the central spin model.  
 
 A third behavior is found in the Ising model on a ring, where the number of dark states grows exponentially with $N$. Here, the dark state condition requires the two neighbors of the monitored spin to be opposite, which leaves $2^{N-2}$ allowed configurations among the 
$N-1$ remaining spins. As a result, $\overline{n}=3/2$ is a constant for all $N$, and the upper bound $ \overline{n} \le 2$ is not saturated in the thermodynamic limit.
 
To conclude, recurrence time in the thermodynamic limit exhibits diverse behaviors. In most systems, the upper bound is saturated, yielding $\bar{n} = 2$. However, the approach to this limit is crucial. As system size increases, the number of dark states may remain constant or grow rapidly with $N$, leading to different paths to the asymptotic limit [Eqs. (\ref{eqN111}) and (\ref{eqN222})]. For the Ising model, the number of dark states grows exponentially, resulting in $\bar{n} < 2$, even in the thermodynamic limit.

\section{Finite Number of Measurements}

 As mentioned, in the thermodynamic limit $\overline{n}$ reaches a constant value. Given a macroscopic system, will we always record this $\overline{n}$ in the $N\rightarrow \infty$ limit, or can we explore various fractional ratios for $\overline{n}$? We claim that the latter is certainly possible, using finite-time experiments. 

The finite number of measurements already played a role in the results we presented above. For example, in Fig. \ref{fig1}(A,B), the resonances of the mean recurrence time are broadened; this is related to the fact when a new dark state is formed, there exists an eigenvalue of the survival operator with $|\xi|\rightarrow 1$, hence the number of measurements required for convergence diverges, leading to the broadening of resonances. These effects are also very evident in the simulation results with the quantum computer, presented below. Here, we further take advantage of the finite number of measurements to measure the coupling magnitude in the spins.  In physical systems, a monitored spin is typically coupled strongly to spins in its vicinity, but weakly to other spins, because the coupling is distance-dependent. We find that using a finite number of measurements, the measured $\overline{n}$ is indifferent to weakly coupled spins. In other words, when the number of measurements is not very large, the system exhibits the behaviors of a small $N$ system. This enables the sensing of different coupling magnitudes and characterizing the distribution of bath spins via the measurement of $\overline{n}$.

As an illustration, we consider the XXX central spin model (see SI Appendix, section 6 for details). We assign various $g_i$ values and compute the conditional recurrence time $\overline{n} = \sum_{n=1}^{K} n F_n/\sum_{n=1}^{K} F_n$ for different $K$. $K$ represents the finite number of measurements, controlled in the experiment. In Fig. \ref{fign3}(A), $g_1 = 1$ and $g_2 = 0.01$. When $K \to \infty$, the theory presented in the previous section is valid. When $10^1 < K < 10^4$, the weakly coupled spin yields no contribution to $\overline{n}$ and $\overline{n} = 3/2$, effectively making it a two-spin system, with a substantial plateau in $\overline{n}$. As $K$ increases, $\overline{n}$ shifts from $3/2$ to $7/4$, where now $\overline{n}$ can ``sense" the second weakly coupled bath spin. The color demarcation corresponds to the approximate transition point located at $K \approxeq -\log2/\log(|\xi|)$. We next examine a four-spin system, with $g_i = \{1,0.05,0.0025\}$. Depending on the magnitudes of the couplings, $\overline{n}$ senses effectively first one bath spin, then two bath spins, and finally three,  as $K$ increases. Consequently, the magnitude of the coupling correlates with the number of measurements required to observe the influence of the weakly coupled spins on $\overline{n}$.

\section{Integer Mean Recurrence Time}

 Already in Fig. \ref{fig1} we demonstrated that, for certain initial conditions,  $\langle n \rangle$ is an integer. This phenomenon is related to a special topological effect, where the many-body dynamics can be mapped onto a single particle problem. For the Heisenberg chain model, if we start in state $\ket{\uparrow,\downarrow,\downarrow,\cdots}$, we find $ \langle n \rangle = N$, namely, an integer mean recurrence time. The examples of $N=4$ and 6 were presented already in Fig. \ref{fig1}. Thus, for this initial state and $N \gg 1$, $\langle n \rangle$ is much larger than the ensemble $\overline{n}$ in Eq. (\ref{eqTB}). Hence the state with integer quantized $\langle n \rangle$ needs far more measurements to be detected. 

Interestingly, we find that the integer $\langle n \rangle$ results from the strong correlation between the bath and the measured spin. Namely, once the measured spin is recorded in state ``up", it implies all other spins are ``down", giving complete knowledge of the bath's state. This is because the dynamics can be described with an effective single quasi-particle picture: a single spin excitation. We found, with this initial condition, that the dynamics are limited to states with single-spin excitations. This means that along the measurement-induced evolution, the effective Hilbert space is composed of $N$ single spin excitation states, instead of $2^N$ states in general. The measurement of the spin ``up" yields full information on the system, since once we record the monitored spin in state ``up", the other bath spins must be in state ``down". Hence the measurement is effectively a full-space measurement, and so equivalent to a single particle evolving among $N$ states. In this case, theory predicts that the amplitude of the first detection time, or more specifically, the generating function of the first detection amplitude, gains a phase winding in the complex plane~\cite{Grnbaum2013,Yin2019}. Using the full space measurement theory, we get an integer mean recurrence time, which is $N$. This implies that special initial states can have profound differences in their topological properties and hence in their recurrence, compared with typical initial conditions.

\section{Universality of Upper Bounds in Spin-X Systems}

Building on our previous analysis of spin-\(\frac{1}{2}\) models, we extend our investigation to spin-1 chains to confirm that the mean recurrence time is fractionally quantized and to examine the universality of the upper bound [Eq. (\ref{eqTB})]. The model we used here is the Heisenberg model given in Eq. (\ref{eqN101}) with $S_i^\alpha$ replaced by the spin-1 operator. 

Leveraging the additional degrees of freedom present in the spin-1 model, we can define two types of measurements: (1) a measurement that precisely determines the spin state as +1, 0, or -1, which we refer to as the full-knowledge measurement, and (2) a measurement that only ascertains whether the spin is in the +1 state or not, without distinguishing between 0 and -1. We denote this as the partial-knowledge measurement. For both cases, the mean recurrence time is fractionally quantized (see SI Appendix, section 7 and Fig. S5 for details). Notably, the values differ between the two measurement protocols, reflecting different effective Hilbert spaces arising from the distinct measurement schemes. Excluding resonances, we see that for full-knowledge measurement, the $\overline{n}$ is 7/3, while for partial-knowledge measurement, it is 23/9. Thus, the value of the fractional number depends on the amount of information gained in each measurement.

We can derive a general upper bound for the recurrence time in the case of the partial-knowledge measurement, following the approach used for the spin-\(\frac{1}{2}\) case. As in the spin-\(\frac{1}{2}\) case, we define two measurement operators: $D^{+1} =\sum_{b} \ket{+1,b}\bra{b, +1}$ and its complement $D_{0,-1} =1-\sum_{b} \ket{+1,b}\bra{b, +1}$. Using Eq. (\ref{eq06}), we define the survival operator for spin-1 case as $S=D_{0,-1}U$. The corresponding eigenvalues $\xi$ satisfy ${\rm det} (S - \xi)=0$. Following Eq. (\ref{eq09}), the ensemble mean recurrence time for spin-1 model under partial-knowledge measurement is:
\begin{equation}
	\overline{n} = {  \mbox{number of} \  \xi_j \ \mbox{in the unit circle} \over 3^{N-1} } \leq 3.
	\label{eqn20}
\end{equation}
The result for the upper bound of the mean recurrence time in Eq. (\ref{eqn20}) can be interpreted as a direct consequence of the spin system's local degrees of freedom. Specifically, the upper bound reflects how the number of accessible spin states influences the recurrence dynamics.

For any spin-X model under partial-knowledge measurement, where the measurement only reveals whether the spin is in the +X state (without distinguishing between other possible states), the system's recurrence time is quantized in a fractional manner. The upper bound is determined by the number of independent states that the monitored spin can occupy. Thus, for a spin-X system, the recurrence time is bounded by:
\begin{equation}
	\overline{n} \leq 2X+1.
\end{equation} 
which suggests that the recurrence time grows linearly with the spin dimension X. This result highlights the interplay between the system's local spin structure and the measurement process, providing insights into how the complexity of the recurrence dynamics scales with the spin dimension.

\section{Quantum Computer Demonstration }

\begin{figure*}
\centering
\includegraphics[width=2\columnwidth, trim=0cm 0cm 0cm 0cm, clip]{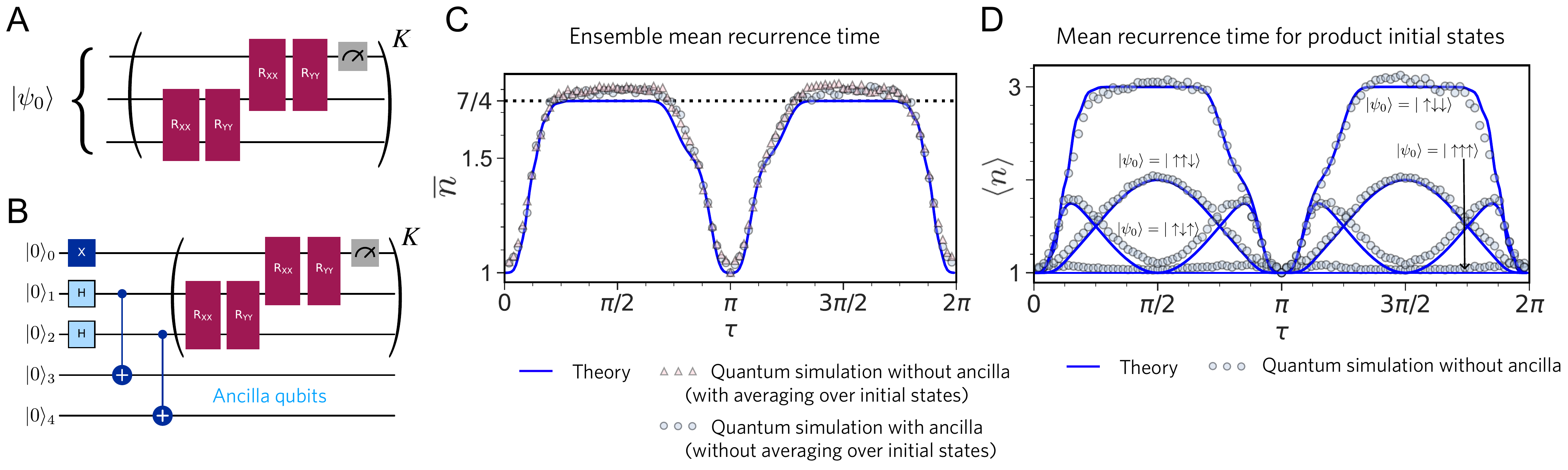}
\caption{Observation of recurrence times on a quantum computer. (A) Quantum circuit implementing repeated measurements. We consider four initial states: $\ket{\uparrow \downarrow \downarrow},  \ket{\uparrow \downarrow \uparrow},  \ket{\uparrow\uparrow \downarrow}, \ket{\uparrow\uparrow\uparrow}.$ For each state, we compute the mean \(\langle n \rangle\) and then average these values to obtain \(\bar{n}\).  (B) Quantum circuit implementing repeated measurements for the entangled initial state with ancilla qubits. The initial state  \(\ket{\psi_{0,\text{av}}}\) is prepared, where qubit~1 is entangled with ancilla qubit~3, and qubit~2 is entangled with ancilla qubit~4. Qubits~3 and~4 serve as ancillas. This configuration effectively averages over all four initial conditions for the recurrence time. (C) Experimental observation of the ensemble mean recurrence time \(\bar{n}\) for three spins with \(K = 20\). The solid line represents the theoretical simulation.  Triangles denote the results obtained by averaging the direct measurements for the four individual initial states shown in (D).  Circles represent the measurement results from the ancilla-based approach described in (B).  (D) The mean recurrence time \(\langle n \rangle\)  is shown for the product initial states  \(\ket{\psi_0} = \ket{\uparrow \downarrow \downarrow}, \ket{\uparrow\uparrow \downarrow},  \ket{\uparrow \downarrow \uparrow}, \ket{\uparrow \uparrow \uparrow}\), respectively. Averaging the result in (D) gives the mean $\bar{n}$ in (C).}
\label{fig:XYN20}
\end{figure*}

We now investigate the fractional quantization of recurrence time using the ibmq\_sherbrooke quantum computer. The first step is implementing the unitary evolution of the Heisenberg spin model with $H=\sum_{\langle i, i+1\rangle } ( S_{i}^{x} S_{i+1}^{x} + S_{i}^{y} S_{i+1}^{y})$ using first order Trotterization (\textit{Materials and Methods}). Such a model is also called the kicked XX model \cite{PRXQuantum.5.010308}. The system under examination consists of a linear chain of three spin-\(\frac{1}{2}\) particles (three qubits, \(N = 3\)) with nearest-neighbor interactions and open boundary conditions. The recurrence of the leftmost spin [equivalently, qubit 0 in Fig. \ref{fig:XYN20}(A)] is measured every \(\tau\) time unit. In this setup, the theory predicts $p = 7$ and $q = 4$, except at specific resonance values of $\tau$. To arrive at this conclusion using Eq. (\ref{eq11}), we note that the system generically supports a unique dark state, corresponding to the configuration in which all qubits (or spins) are in the ``down" state. In the quantum simulation, the unitary evolution operator is \( U_{\text{Trotter}} \), see details in \textit{Materials and Methods}, which is an approximation of the exact unitary evolution of the spin model. A natural question is, therefore, how well this approximation performs in the context of fractional recurrence times. In particular, for large \( \tau \), the two unitaries are clearly non-identical, so one might naively expect deviations to be significant for large \( \tau \). However, our analysis shows that the number of dark states remains unchanged between the exact unitary $U=\exp(-i H \tau)$ and its Trotterized counterpart \( U_{\text{Trotter}} \), preserving the key mechanism behind fractional mean recurrence times. This suggests that Trotterization does not obscure the effect but rather provides an experimentally feasible way to verify it. The robustness of fractional recurrence against Trotterization errors further underscores its generality and relevance for quantum simulations.

We implement the quantum circuits in Fig.~\ref{fig:XYN20}(A) on the quantum computer. 
As an error prevention scheme, we use dynamical decoupling, which is designed to maintain quantum coherence by counteracting environmental noise and decay during the measurement \cite{PhysRevApplied.20.064027}. This is achieved by applying a sequence of control pulses to the quantum system, effectively neutralizing unwanted interactions with the environment. We calculate the mean recurrence time for various initial states by evaluating $F_n$, the probability of measuring the qubit 0 the first time in the state $\ket{\uparrow}$ after $n$ measurements. The mean recurrence time is determined using the formula:
$
\langle n \rangle = \sum_{n=1}^{K} n F_n/\sum_{n=1}^{K} F_n,
$
where $K$ represents the total number of measurements. Our results confirm the fractional quantization of recurrence time predicted theoretically, as shown in Fig.~\ref{fig:XYN20}(C), triangles.  The measured recurrence times for the four different product initial states are illustrated in Fig.~\ref{fig:XYN20}(D); by averaging these initial states, we obtain the $\bar{n}$ in Fig.~\ref{fig:XYN20}(C). These findings successfully validate the presence of fractional recurrence times when using a quantum computer.

The resonances in Fig.~\ref{fig:XYN20}(C) are broader than those shown in Fig.~\ref{fig1}. This broadening arises from the use of \( K = 20 \) measurements. While increasing \( K \) can, in principle, sharpen the resonances, it also amplifies the impact of noise. Therefore, \( K \) is chosen to be small to mitigate noise effects. A more detailed analysis of resonance broadening remains an open challenge.  This poses an issue if the system size is large; hence, we tackle the averaging procedure with a new approach.

\subsection{Using entanglement to measure fractional quantization} As mentioned, averaging over all bath states can be challenging for large systems, as we have $2^{N-1}$ bath states \cite{PhysRevLett.95.140501}. Instead, we use a unique initial state to record fractional quantization. We now explain this method in detail, limiting ourselves to the $N=3$ system, as done in the quantum simulation.  The idea is to introduce two ancilla qubits (qubits 3 and 4), where qubit 1 is entangled with ancilla qubit 3, and qubit 2 is entangled with ancilla qubit 4 [see Fig. \ref{fig:XYN20}(B)]. By doing so, we generate a pure initial state in the quantum computer, represented as $\ket{\psi_{0,\mathrm{av}}}=\ket{\uparrow}\otimes \!\!\sum_{i,j\in\{\downarrow,\uparrow\}}\!\ket{ij}\otimes\ket{ij}_{\mathrm{anc}}/2$. Here the subscript “anc” denotes the ancilla qubits, and $\ket{ij}_{\mathrm{anc}}$ is the corresponding two-qubit ancilla basis state, with $i,j \in \{\downarrow, \uparrow\}$. Here, the first $\ket{ \uparrow }$ represents the recorded spin or qubit, which is in the state ``up" to start with. As usual,  the unitary is acting only on the first three qubits, and not on the ancillas. By tracing out the two ancilla qubits, we obtain $\text{Tr}_{3,4}\left( |\psi_{0,\text{av}}\rangle \langle \psi_{0,\text{av}}| \right) 
=\, \ket{\uparrow} \bra{\uparrow} \otimes \frac{1}{4} \big( 
\ket{\downarrow\downarrow} \bra{\downarrow\downarrow} 
+ \ket{\downarrow\uparrow} \bra{\downarrow\uparrow} 
+ \ket{\uparrow\downarrow} \bra{\uparrow\downarrow} 
+ \ket{\uparrow\uparrow} \bra{\uparrow\uparrow} \big)$.
 We see that now we have four recurrence states $\ket{\uparrow \downarrow \downarrow}, \ket{\uparrow \downarrow \uparrow}, \ket{\uparrow \uparrow \downarrow}, \ket{\uparrow \uparrow \uparrow}$, with equal probability $1/4$.
 The recurrence time for the single state $|\psi_{0,\text{av}}\rangle$ thus corresponds to the ensemble recurrence time for the three-qubit case. Therefore, instead of averaging over all initial states, by utilizing entanglement and ancilla qubits, we can experimentally measure ensemble recurrence times by preparing a single initial state. This method can be generalized to an $N$-qubit system, requiring $N-1$ ancilla qubits. This approach eliminates the need to average over an exponentially large number of initial states, making it significantly easier to measure ensemble recurrence times even in large many-body systems.

How to prepare this state for a bath with two qubits is described in \textit{Materials and Methods} and displayed in Fig.~\ref{fig:XYN20}(B).  We apply a Hadamard gate on each bath qubit followed by a CNOT gate, which entangles each of the bath qubits with an ancilla bath qubit: 
\begin{equation}
\begin{aligned}
&\fbox{H} \ket{\downarrow} \, \fbox{H} \ket{\downarrow} \otimes \ket{\downarrow\downarrow}_{\text{anc}} \rightarrow \frac{\ket{\downarrow} + \ket{\uparrow}}{\sqrt{2}} \cdot \frac{\ket{\downarrow} + \ket{\uparrow}}{\sqrt{2}} \otimes \ket{\downarrow\downarrow}_{\text{anc}} \\
&\xrightarrow{\text{CNOT}} \frac{1}{2} \!\!\sum_{i,j\in\{\downarrow,\uparrow\}}\!\ket{ij}\otimes\ket{ij}_{\mathrm{anc}}
\end{aligned}
\label{eq:evolution}
\end{equation}
Hence, we obtained the desired state $|\psi_{0,\text{av}}\rangle$.
In Fig.~\ref{fig:XYN20}(C) [triangles], we present the average $\bar{n}$ for the four initial states $\ket{\uparrow \downarrow \downarrow}$, $\ket{\uparrow\uparrow\downarrow}$, $\ket{\uparrow \downarrow\uparrow}$, and $\ket{\uparrow\uparrow\uparrow}$. We compare the results by measuring the $\bar{n}$ with the entangled state (circles). Both averaging methods yield the same result $\bar{n} \approx 7/4$ [Fig.~\ref{fig:XYN20}(C)], except at the broadened resonances. We get the integer return time $\langle n \rangle = 3$ for the initial state $\ket{ \uparrow \downarrow \downarrow}$ [Fig.~\ref{fig:XYN20}(D)], which is expected from the above theory, since this is the integer quantized recurrence time in this example. Other values of $\langle n \rangle$ for other initial states are also shown in Fig.~\ref{fig:XYN20}(D). 

The quantum simulation results closely match the theoretical predictions, with only a deviation of 5\%, confirming the first on-device observation of fractional recurrence times, providing strong validation of the theory. For larger spin models or more complex circuits, it will be necessary to incorporate a noise model and employ more elaborate error mitigation techniques to effectively reduce the impact of noise, which can be left for future work. Another challenge is to characterize the width of the resonances and their dependence on \(K\). We expect that the width of resonances will diminish as \(K\) increases, at least up to some $K$ where the noise begins to dominate; however, this challenge has not yet been met on a quantum computer.

\section{Summary and Discussion}

We studied the recurrence problem in quantum many-body systems, where a single spin is repeatedly measured at intervals $\tau$ to detect its recurrence to the initial state. Our results demonstrate that the mean recurrence time $\bar{n}$ is fractionally quantized, taking values of the form $p/q$. By analyzing various spin models, including Heisenberg spin chains, Central spin models, and Ising chains, we established that this fractional character is robust and universal, governed by rigid upper and lower bounds. Crucially, we identified a fundamental relationship between the recurrence time and the number of dark states in the system, providing a novel method to probe Hilbert space fragmentation. These theoretical predictions were corroborated by demonstrations on an IBM quantum processor using mid-circuit measurements, confirming that the topological nature of the fractional quantization offers significant resilience against experimental noise. These findings provide a foundation for future studies of measurement-induced phenomena in interacting quantum systems and their application in quantum information science. Below, we discuss the central implications of these findings.

\textbf{1. Universality and Speed Limits of Information Retrieval.} Intriguingly, we derived a universal bound for the recurrence time, $1 \le \bar{n} \le 2$, valid for any spin-1/2 system under measurements. This result interprets the recurrence time as a fundamental timescale for information retrieval: it sets both minimum and maximum ``speed limits" for how quickly a quantum state can complete a cycle and be measured by subspace detection. The upper bound acts as a fundamental constraint, independent of the interaction details or system size. We further extended this analysis to spin-$X$ systems, finding that the upper bound scales linearly with the spin dimension ($\bar{n} \le 2X+1$). This confirms the universality of the bounds and highlights the intrinsic connection between the local degrees of freedom and the complexity of the recurrence dynamics.

\textbf{2. Probing Hilbert Space Fragmentation via Dark States.} Crucially, the fractional recurrence time is determined by the number of dark states in the system. Dark states are initial conditions that are never detected; their presence indicates the fragmentation of Hilbert space and ergodicity breaking. We derived the relation $\bar{n} = 2 - N_{dark}/2^{N-1}$, where $N_{dark}$ denotes the number of dark states. This allows the recurrence time to serve as a diagnostic tool. In systems where the dark space is substantial, the bright subspace contracts, leading to lower recurrence times.
This is analogous to searching for a lost key in a room where part of the room is walled off, thereby making the search (or recurrence) faster. Thus, by measuring the recurrence time, one can infer the dimension of the dark subspace, providing a novel method to probe ergodicity breaking without requiring full state tomography.

\textbf{3. Scaling and the Thermodynamic Limit.} The behavior of quantum systems as they approach the macroscopic limit is of interest. We identified five distinct scaling behaviors for $\bar{n}$ in the thermodynamic limit ($N \to \infty$): (1) $\overline{n}$ is $N$ independent as found in the Ising model, $\overline{n}=1$ for $Z$ direction measurement and $\overline{n}=2$ for measurement along $X$ and $Y$. This difference is controlled by the commutation relation between measurement operator and Hamiltonian. The Ising Hamiltonian is diagonal in the $Z$ basis and hence commutes with $Z$ direction measurement. Therefore, any states with the monitored spin in “down” state remain dark under the repeated measurements, giving $N_{dark} = 2^{N-1}$ and $\bar{n}=1$. By contrast, measurements in $X/Y$ basis do not commute with $H$. A state prepared in $X$ or $Y$ basis is a superposition of $Z$ basis states, which acquire different phases during the unitary evolution and develop overlap with the detected state. The dark manifold is then absent, so $\bar{n}=2$. (2) $\overline{n}$ exponentially approaches upper bound 2 in the Heisenberg spin model; (3) $\overline{n}$ approaches 2 as a power-law in the central spin model; (4) $\overline{n}$ is 2 for the strongly anisotropic DM case, as there is no dark states in the Hilbert space.  (5) $\overline{n}$ is 3/2 for the Ising ring model, and hence in this case the upper and lower bounds are not saturated even in the thermodynamic limit. The speed of approach to saturation is determined by the scaling of the number of dark states with system size $N$, which acts as a proxy for Hilbert space fragmentation. In locally interacting systems like the Heisenberg chain, the dark state fraction vanishes exponentially ($\sim 2^{-N}$), leading to rapid saturation and signifying robust ergodicity. Conversely, systems with high symmetry, such as the Central spin model, exhibit a slower power-law approach ($2-\overline{n} \sim N^{-1/2}$) due to an algebraically scaling dark subspace, reflecting weak ergodicity breaking. Finally, in systems like the Ising model, the dark subspace scales proportionally to the Hilbert space, preventing saturation entirely.

\textbf{4. Measurement-Induced Resonances.} While fractional quantization is topologically protected, we observed that the integers $p$ and $q$ undergo abrupt transitions at specific ``resonant" values of the time interval $\tau$. These resonances occur when new dark states form due to the interference between the measurement periodicity and the system's energy gaps. This sensitivity allows the recurrence time to act as a spectroscopic tool: by scanning $\tau$ and observing the drops in $\bar{n}$, one can detect the underlying periodicities of many body systems that yield dark states under monitoring.

\textbf{5. Emergence of Integer Quantization.}  For specific initial states, we showed that the complex many-body system can be effectively mapped onto a dynamical system with a single spin excitation, leading to the presence of integer quantized recurrence times, see for example  Fig. \ref{fig:XYN20}(D), where the integer three is recorded experimentally.  This is an example of how subspace measurements on a single spin can effectively behave as full space measurements. This happens due to a strong correlation between the measured spin and the bath spins. When the system is large, the result is a large mean recurrence time, equal to $N \tau$, if compared to typical initial conditions where $ \overline{n} \leq 2$.

\textbf{6. Experimental Robustness, Entanglement, and Quantum Benchmarking.} Finally, our implementation on the IBM quantum computer addresses the practicality of these invariants in the noisy intermediate-scale quantum era. We successfully observed the predicted fractional quantization and resonances, validating the theory. Despite the number of qubits being small, the experimental setup involves a relatively deep circuit with mid-circuit measurements, capturing nontrivial quantum dynamics beyond simple toy models. This complexity highlights that meaningful recurrence behavior can still be probed within constrained quantum resources. More profoundly, the recurrence time in our theory is defined as an average over non-monitored qubit states, which introduces an inherent averaging step. To address this, we propose a method to bypass the need for such averaging by exploiting entanglement and introducing ancilla qubits into the system. This approach allows us to measure the fractional recurrence time directly without relying on state averages. Furthermore, we demonstrate the robustness of the fractional quantization: both the unitary dynamics of the Heisenberg XX model we use as an example and its Trotterized version yield consistent $p/q$ values, despite the differences in their implementations. These results suggest that the topological invariant $p/q$ can function as a robust benchmark for noisy quantum processors with mid-circuit measurement capabilities. We demonstrate that the number of ancillas increases linearly with the number of qubits, whereas direct measurements of fractional quantization \textemdash\ or equivalently, of the number of dark states \textemdash\ require an exponentially large number of operations. This reveals a quantum advantage and speedup in the evaluation of the number of dark states in many-body systems.

\section*{Materials and Methods}
\subsection*{Numerical Simulation}
The plots are prepared by directly simulating the subspace measurements process based on Eq. (\ref{eq01}). We initially construct the Hamiltonian $H$ in the Fock space for a Heisenberg spin chain, central spin models, and Ising model comprising $N$ spins. The product recurrence initial state $\ket{\uparrow, \downarrow,\downarrow, \cdots, \downarrow}$ is then represented as a vector of dimension $2^N$. With the initial state and $H$, we numerically compute the first detection amplitude $\phi_1$. The first detection probability is given by $F_1= ||\phi_1||^2$, which is recorded. We then proceed to calculate $F_2$. In the first step, the failed measurement projects out the state with the monitored spin ``up", see Eq. (\ref{eq01}). This is achieved by setting the state that overlaps with the monitored spin ``up" to zero, effectively mimicking the back-action of projection $1- D^{\uparrow}$. The measured state becomes the new initial state for the calculation of $F_2$. Similar to the calculation of $F_1$, we allow the system to evolve for time $\tau$ via $U(\tau)$, then compute $\phi_2$. The first detection probability $F_2 = || \phi_2||^2$. This procedure is repeated to numerically calculate $F_3, F_4,\cdots, F_n$. The first detection recurrence time is given by $ \langle n \rangle = \sum_{n} n F_n$.

\subsection*{Quantum Computer Implementation}
All experiments were performed on the ibmq\_sherbrooke superconducting quantum processor. The kicked XX dynamics for three qubits with open boundaries were realized via a first-order Trotter step of duration \(\tau\), $U_{\mathrm{Trotter}}(\tau)=\prod_{\langle i, j \rangle}\exp\left(-i\frac{1}{2} S_i^xS_{j}^x \tau\right)\exp\left(-i \frac{1}{2} S_i^yS_{j}^y \tau\right)=\prod_{\langle i,j\rangle} R_{X_iX_j}(-\tau)\,R_{Y_iY_j}(-\tau)$,
where \(R_{X_iX_j}\) and \(R_{Y_iY_j}\) are two-qubit rotations compiled to the device’s native cross-resonance entangling gates and single-qubit rotations; qubit placement and routing were chosen to minimize SWAPs [Fig.~\ref{fig:XYN20}(A,B)]. Typical readout assignment error was \(\sim 5\times10^{-3}\); two-qubit gate error ranged from \(5\times10^{-3}\) to \(1\times10^{-2}\). Dynamical decoupling (periodic \(\pi\)-pulse sequences) was applied to idle qubits during inter-measurement windows to suppress dephasing and low-frequency noise. For each \(\tau\), qubit~0 was measured projectively after each application of \(U_{\mathrm{Trotter}}(\tau)\). On the ibmq\_sherbrooke backend, the maximum number of repeated measurement shots was limited to 20 by backend-specific software constraints; all experiments were configured to respect this cap.
 The first-hit distribution \(F_n\) over \(K=20\) measurement cycles was collected to compute the mean recurrence time \(\langle n\rangle\).

Ensemble averaging over bath configurations was implemented in two ways. \emph{Direct averaging}: the system was initialized in each of the four product states \(\{\ket{\uparrow\downarrow\downarrow},\ket{\uparrow\downarrow\uparrow},\ket{\uparrow\uparrow\downarrow},\ket{\uparrow\uparrow\uparrow}\}\); the corresponding \(\langle n\rangle\) values were measured and averaged to obtain \(\bar n\) [Fig.~\ref{fig:XYN20}(C,D)]. \emph{Ancilla-assisted averaging}: two ancilla qubits (3 and 4) were introduced as record registers [Fig.~\ref{fig:XYN20}(B)]. Bath qubits 1 and 2 were prepared in \((\ket{\downarrow}+\ket{\uparrow})/\sqrt{2}\) via Hadamards and then entangled with ancillas via CNOTs to yield $\ket{\psi_{0,\mathrm{av}}}=\ket{\uparrow}\otimes \tfrac{1}{2}\!\!\sum_{i,j\in\{\downarrow,\uparrow\}}\!\ket{ij}\otimes\ket{ij}_{\mathrm{anc}}$. Only the system qubits underwent \(U_{\mathrm{Trotter}}(\tau)\); ancillas were left idle and disregarded at readout, effectively tracing them out and producing a uniform mixture over the four product states. The \(\bar n\) extracted from this single-state protocol agreed with the direct average within experimental uncertainty. Resonance features were broader than ideal due to finite \(K\); increasing \(K\) sharpens features but amplifies noise, so \(K=20\) was adopted as a practical compromise.

\section*{Data, Materials, and Software Availability}
The experiment data supporting the findings of this study are available at Zenodo and can be accessed via DOI: \href{https://doi.org/10.5281/zenodo.18429670}{https://doi.org/10.5281/zenodo.18429670}~\cite{liu_2026_zenodo}.

\section*{Acknowledgments}
The support of Israel Science Foundation's Grant No. 2311/25 is acknowledged. Q.L. was also supported by the National Natural Science Foundation of China (Grant No. 12447113). We acknowledge the use of IBM Quantum services for this work. The views expressed are those of the authors, and do not reflect the official policy or position of IBM or the IBM Quantum team. In this paper we used ibmq\_sherbrooke, which is one of the IBM Quantum Eagle processors.

\bibliography{ref}

\appendix

\clearpage

\onecolumngrid
\appendix 

\newcommand{\supplementarytitle}{
    \begin{center}
        \Large\bfseries Supplementary Information for \\ 
``Fractionally Quantized Recurrence Detection Times in Monitored Quantum Many-Body Systems''
    \end{center}
    \vspace{0.10cm} 
}

\supplementarytitle

\section{The recurrence time with superposition initial states}
 In the main text, we considered mainly initial states, which are product states. In this section, we consider the recurrence time with superposition initial states. As the example studied in Figs. 1(a,b) in the main text, we consider a system with four spins. The superposition initial states can be written as:
\begin{equation}
\begin{aligned}
	&\vert \psi_{1,2} \rangle = \frac{ \vert \uparrow \uparrow \uparrow \uparrow \rangle \pm \vert \uparrow \uparrow \uparrow \downarrow \rangle}{\sqrt{2}}, 
&\vert \psi_{3,4} \rangle = \frac{\vert \uparrow \uparrow \downarrow \uparrow \rangle \pm \vert \uparrow \uparrow \downarrow \downarrow \rangle}{\sqrt{2}},  \\ 
&\vert \psi_{5,6} \rangle = \frac{\vert \uparrow \downarrow \uparrow \uparrow \rangle \pm \vert \uparrow \downarrow \uparrow \downarrow \rangle}{\sqrt{2}}, 
&\vert \psi_{7,8} \rangle = \frac{\vert \uparrow \downarrow \downarrow \uparrow \rangle \pm \vert \uparrow \downarrow \downarrow \downarrow \rangle}{\sqrt{2}}.
\end{aligned}
\label{eq33ap}
\end{equation}
With these initial conditions, we first find $\langle n \rangle$ and then we average to find the ensemble mean $\bar{n}$. Here the recurrence times $\langle n \rangle$ for $\vert \psi_1 \rangle$ and $|\psi_2\rangle$ are the same; similarly with the pairs $|\psi_3\rangle$ and $|\psi_4\rangle$, $|\psi_5\rangle$ and $|\psi_6\rangle$, and $|\psi_7\rangle$ and $|\psi_8\rangle$. Hence, we see four plots for eight initial states. Comparing Fig. 1(A,C) in the main text, we found that the ensemble mean recurrence time denoted $\bar{n}$ remains the same as that of the product initial states. Thus, as guaranteed by the recurrence theorem in the main text, the fractional quantization of the ensemble mean recurrence time is unaffected by the choice of basis.

\section{Heisenberg spin chain with Dzyaloshinskii-Moriya interaction}
\label{apxo}

\begin{figure}[!htbp]
    \centering
    \includegraphics[width=0.6\columnwidth]{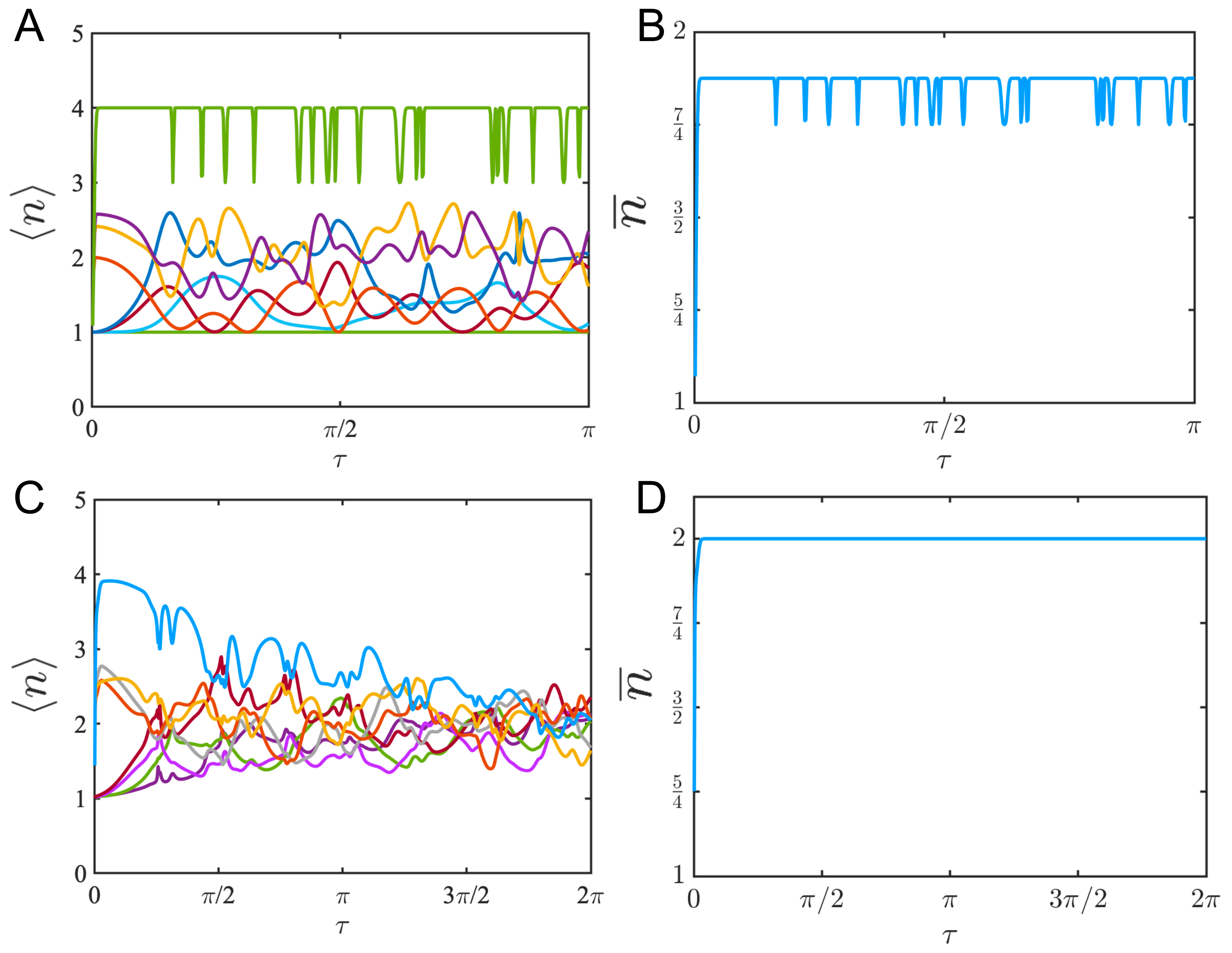}
    \caption{Mean recurrence time for the Heisenberg spin chain with a Dzyaloshinskii-Moriya interaction for a system with four spins. (A) $\langle n \rangle$ for product initial states with four spins. Here we choose $D_{12} = \{2, 0, 0\},  D_{23} = \{1.5, 0, 0\},  D_{34} = \{0.9, 0, 0\}.$ (B) The corresponding ensemble mean $\overline{n}$ versus sampling time $\tau$. (C) The same as (A) but with increasing anisotropy, where  $D_{12} = \{0.4, 0.1, 0.23\},  D_{23} = \{0.1, 0.1, 0.8\},  D_{34} = \{0.1, 0.1, 0.1\}.$ (D) The corresponding ensemble mean versus sampling time $\tau$. Due to the anisotropic nature of \( H \), the system has no generic dark states, causing \( \bar{n} = 2 \).  In the Zeno limit, \( \bar{n} \) exhibits a discontinuity: it is \( 1 \) at \( \tau = 0 \) but jumps to \( 2 \) for any finite \( \tau \). }    
    \label{fig3newadd}
\end{figure}

In this section, we examine the case of the Heisenberg Hamiltonian incorporating the Dzyaloshinskii-Moriya(DM) interaction. The DM interaction introduces antisymmetric exchange terms that distort spin configurations. The full Hamiltonian of the system can be written as $H=H_{Hei}+H_{DM}$, where $H_{Hei}$ is the Heisenberg spin-chain Hamiltonian defined in Eq. (1) in the main text. The newly added DM interaction is given by:
\begin{equation}
    H_{DM} = \sum_{\langle i, i+1 \rangle} \vec{D}_{DM} \cdot (\vec{S}_i \times \vec{S}_{i+1}),
\end{equation}
where \(\vec{D}_{DM}\) is the DM vector, determining the strength and orientation of the interaction. As an illustrative example, Fig.  \ref{fig3newadd} depicts the mean recurrence time for a Heisenberg spin chain with DM interaction, with a system size of \(N = 4\). In the calculation, we use $D_{12} = \{0.4, 0.1, 0.23\},  D_{23} = \{0.1, 0.1, 0.8\},  D_{34} = \{0.1, 0.1, 0.1\}$ for Fig. \ref{fig3newadd}(A,B). We use $D_{12} = \{2, 0, 0\},  D_{23} = \{1.5, 0, 0\},  D_{34} = \{0.9, 0, 0\}$ for Fig. \ref{fig3newadd}(C,D). Its direction is generally dictated by the symmetry of the lattice. The term \(\vec{S}_i \times \vec{S}_{i+1}\) represents the cross product of spin operators.

 As shown in the figure, depending on $\vec{D}_{\textit{DM}}$, the system can either express fractional quantization, as seen in Fig. \ref{fig3newadd}(B), or integer quantization, exemplified by $\bar{n} = 2$ in Fig. \ref{fig3newadd}(D), revealing distinct underlying physical mechanisms. The latter arises in a regime where the Hamiltonian $H$ is highly non-symmetric. As we demonstrate in the main text, this asymmetry leads to a significant suppression in the number of dark states present in the system. Furthermore, we observe that resonances in $\bar{n}$ appear in certain cases [e.g., Fig. \ref{fig3newadd}(B)] but are absent in others [e.g., \ref{fig3newadd}(D)]. This behavior is closely correlated with the degree of symmetry breaking in the Hamiltonian. Despite the additional complexity introduced by the DM interaction, our analysis reveals that the fractional quantization of recurrence times remains intact.

\section{Recurrence time and eigenvalues of the survival operator}
\label{ap1}

\begin{figure}[!htbp]
    \centering
    \includegraphics[width=\columnwidth]{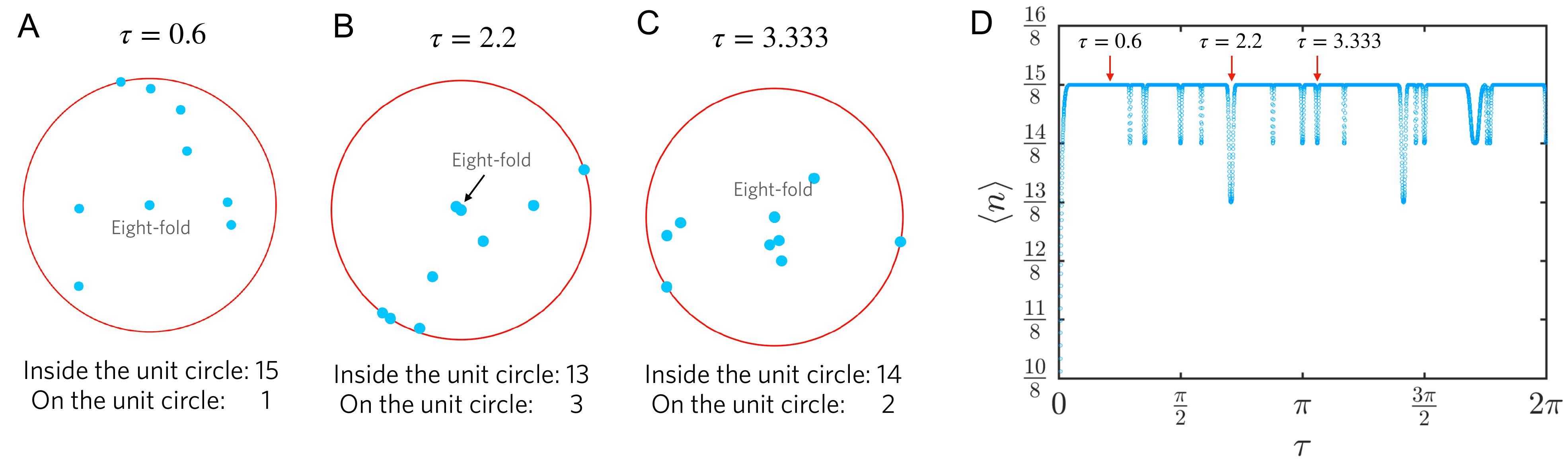}
    \caption{Eigenvalues of the survival operator are distributed on and in the unit circle, fig (A-C), for the four-spin Heisenberg chain. In (D), we show the $\overline{n}$ for the corresponding choices of $\tau$ (see red arrows). (A) For $\tau =0.6$, 15 eigenvalues of the survival operator are located within the unit circle, with one eigenvalue on the unit circle. The ensemble mean recurrence time is determined by the number of eigenvalues within the unit circle divided by $2^{N-1}=8$ since $N=4$. Hence for $\tau = 0.6$ we have $\overline{n}=15/8$. (B) At $\tau \simeq 2.22$, $\overline{n}$ decreases to $13/8$, aligning with the number of eigenstates within the unit circle as shown in (B). (C) When $\tau \simeq 3.333$, $\overline{n}$ jumps to $14/8$, and the count of eigenvalues within the unit circle is 14.}    
    \label{figsub1}
\end{figure}

In this section, we present further details on how $\overline{n}$ is given by the number of eigenvalues of the survival operator. In the main text, the theoretical Eq. (15) gives a perfect match with the numerical $\overline{n}$, as shown in Figs. 2(b,d) in the main text. In Fig. \ref{figsub1}, we plot the exact locations of the eigenvalues of the survival operator for specific sampling times $\tau$ for the four-spin model described in the main text. This offers a more graphic presentation of how the recurrence time is connected to the eigenvalues $\xi$. Specifically, we plot the eigenvalues $\xi$s for different values of $\tau$ for the four-spin Heisenberg spin chain. This model was defined in the preliminary in the main text. When $\tau = 0.6$, there are 15 $\xi$s inside the unit circle. Using Eq. (15), we find $\overline{n} = 15/8$, consistent with the numerical simulation in Fig. \ref{figsub1}(D). In fact, for any $\tau$ where $ \overline{n} =15/8$, 15 $\xi$s are found inside the unit circle. When $\tau$ is at the resonance near 2.22, the $\overline{n}$ jumps to $13/8$. Now, the number of $\xi$ inside the unit circle is 13, as shown in Fig. \ref{figsub1}(B). Similarly, for $\tau$ at the resonance at 10/3, there are 14 $\xi$s inside the unit circle and $\overline{n} = 7/4$.

\section{Dark states and resonances}
\label{ap2}

We have seen resonances in the ensemble mean recurrence time, for example, in Fig. \ref{figsub1}, which give rise to point-wise discontinuities in $\overline{n}$. In the main text, we showed that this is because of the additional dark states, which appear for special sampling times $\tau$. In this section, we present the detailed steps for the construction of additional dark states for the central spin model. 

We first discuss the general properties of the central spin model following Villazon \textit{et al.} \cite{Villazon2020}. As mentioned in the main text, the Hamiltonian of the XX central spin model is 
\begin{equation}
H= w_0 S_0^z +w \sum_{i=1}^{N-1} S_i^z + \sum_{i=1}^{N-1} g_i (S_0^+S_i^-+S_0^-S_i^+),
\label{eq:hce}
\end{equation}
where $S_i^{\pm} = S_i^x \pm i S_i^y$. The spectrum of this model is solved in \cite{Villazon2020}. The eigenstates of $H$ are divided into four groups:  $|D^+\rangle, |{\cal B}(E)\rangle, |{\cal B}(-E)\rangle$, and $| D\rangle$. We soon show why we denote these symbols. The central spin Hamiltonian is diagonalized by these states. The eigenfunction for the states $|{\cal B}(E)\rangle$ and $|{\cal B}(-E)\rangle$ are 
\begin{equation}
	H|{\cal B}(E)\rangle=E|{\cal B}(E)\rangle, \quad H|{\cal B}(-E)\rangle=-E|{\cal B}(-E)\rangle
\end{equation}
Using the Bethe ansatz, in the subspace with $K$ spin excitations atop the vacuum $|0\rangle$ (all the spins are in the state ``down"), we have \cite{Villazon2020}
\begin{equation}
\begin{aligned}
	&|{\cal B}(E)\rangle
	 = \left[G^{+}(0)+\left(E+\frac{\omega_0}{2}\right) S_0^{+}\right]G^+(v_1)\cdots G^+(v_{K-1})|0\rangle,\\
	&|{\cal B}(-E)\rangle 
	 = \left[G^{+}(0)+\left(-E+\frac{\omega_0}{2}\right) S_0^{+}\right]G^+(v_1)\cdots G^+(v_{K-1})|0\rangle,
	\label{eq:96}
\end{aligned}
\end{equation}
where $ G^{+}(v)=\sum_{i=1}^{N-1} g_i/(1-g_i^2 v) S_i^{+}$. The parameters $v_i$ are given by \cite{Villazon2020}:
\begin{equation}
	1+v_a \sum_{i=1}^{N-1} \frac{ g_i^2}{2(1-g_i^2 v_a)}-v_a \sum_{b \neq a}^{K-1} \frac{1}{v_b-v_a}=0,
	\label{eq:bethb}
\end{equation}
for $a=1 \ldots K-1$. The eigenenergy is given by \cite{Villazon2020}
\begin{equation}
	E^2 = \frac{\omega_0^2}{4}+\left(\sum_{i=1}^{N-1}  g_i^2-\sum_{i=1}^{K-1} \frac{2}{v_i}\right).
\end{equation}
 Each solution for $v_1 \ldots v_{N-1}$ leads to two possible solutions for $E$, exhibiting level repulsion.

 For the states $|D\rangle$, we have $H  |D\rangle = -w_0/2 |D\rangle$, which are the dark states we have discussed in the main text. The states $|D^+\rangle$ have the same bath states as $|D\rangle$ but with the monitored spin in the ``up" state, hence its eigenenergy is $w_0/2$ The resonances are given by the states $|{\cal B}(E)\rangle, |{\cal B}(-E)\rangle$ fits the condition we found, hence we now only focus on the states $|{\cal B}(E)\rangle, |{\cal B}(-E)\rangle$. Using Eq. (\ref{eq:96}), we rewrite the states $|{\cal B}(E)\rangle, |{\cal B}(-E)\rangle$ as: 
\begin{equation}
\begin{aligned}
	&|{\cal B}(E)\rangle ={\cal N}\left[ \mid \downarrow\rangle_0 \otimes |\beta^\downarrow \rangle +\left(E+\frac{\omega_0}{2}\right) \mid \uparrow\rangle_0 \otimes |\beta^\uparrow \rangle \right] ,\\
	& |{\cal B}(-E)\rangle = {\cal N}\left[ \mid\downarrow\rangle_0 \otimes |\beta^\downarrow \rangle +\left(-E+\frac{\omega_0}{2}\right) \mid\uparrow\rangle_0 \otimes |\beta^\uparrow \rangle \right],
\end{aligned}
\label{eq:Bright}
\end{equation}
where the states $|\beta^\downarrow \rangle = G^{+}(0)G^+(v_1)\cdots G^+(v_{K-1})|0\rangle$ and $|\beta^\uparrow \rangle = G^+(v_1)\cdots G^+(v_{K-1})|0\rangle$ denote the states of the bath spins and ${\cal N}$ denotes the normalization.

\begin{figure}[!htbp]
    \centering
    \includegraphics[width=0.8\columnwidth]{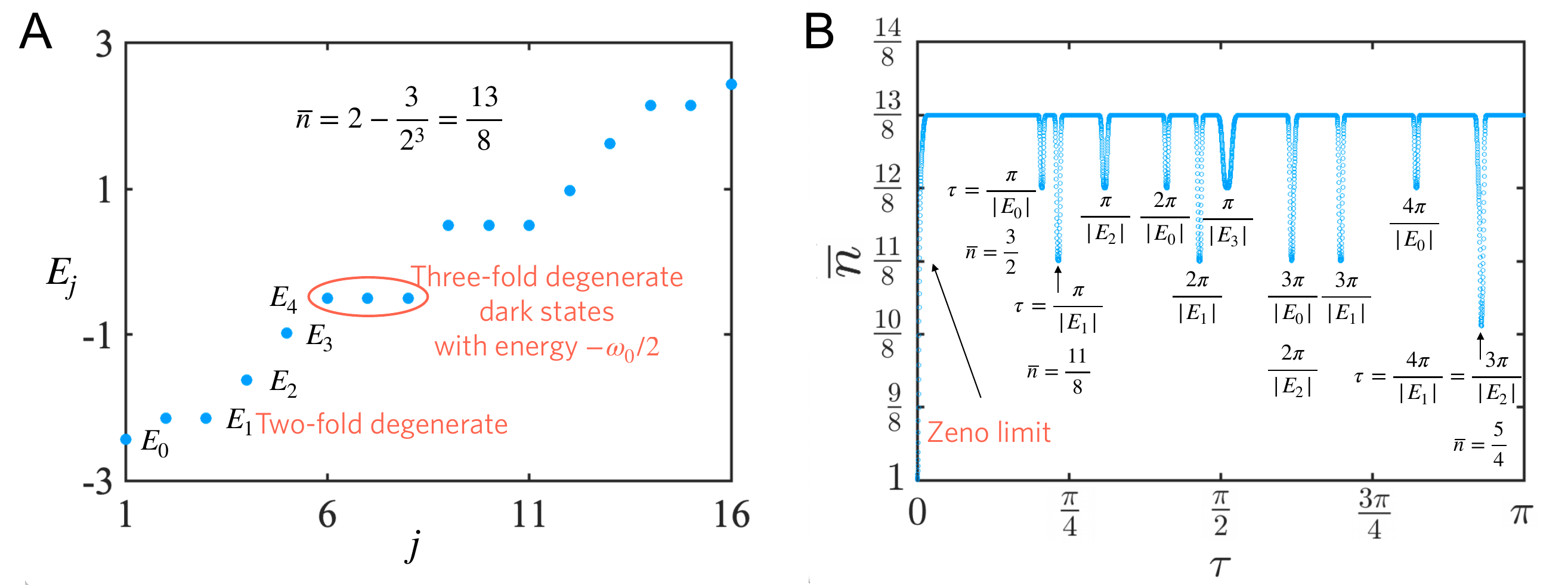}
    \caption{Additional dark states lead to the resonances in $\overline{n}$ for special sampling times. Here $g_1 = 1.221$, $g_2 = 0.864$, and $g_3 = 1.46$, though the main physical effects are not sensitive to these couplings. (A) The energy spectrum of the system. The three states corresponding to the degenerate energy level $E_4= - w_0/2=-1/2$ are typical dark states. The mean recurrence time is determined by the three dark states, hence $\overline{n} = 2-3/8 = 13/8$, aligning with the numerical simulations in (B) except for the resonances. (B) The plot of $\overline{n}$ versus the sampling time $\tau$. When the number of dark states changes, due to the appearance of additional dark states, found for special $\tau$, $\overline{n}$ exhibits the resonances. The energy levels in (A) can be used to understand the locations of dips, namely the special sampling times, as presented in (B).}    
    \label{figsub2}
\end{figure}


We now explain the resonances in the recurrence time. Apart from the dark states that are directly derived from the Hamiltonian's eigenstates ($|D \rangle$), we introduce another category of dark state, which is associated with the sampling time $\tau$ and leads to the resonances in $\overline{n}$, for example, in Fig. \ref{figsub1} and Fig. 2 in the main text. In the central spin model, the interrelation between the eigenstates with energy $E$ and $-E$ [see Eq.(\ref{eq:Bright})] allows for the construction of the additional dark states. The condition $D^{\uparrow} (\alpha_1 |E_1\rangle + \alpha_2 |E_2\rangle) = 0$ in the main text leads to 
\begin{equation} 
\Delta E\tau = 2k \pi, \quad k \quad \text{is an integer.}
\label{eq:darkc}
\end{equation}
 Using Eq. (\ref{eq:Bright}), we find that $U |{\cal B} (E)\rangle = \exp(-i k \pi) |{\cal B} (E)\rangle$ and $ U |{\cal B} (-E)\rangle = \exp( i k \pi) |{\cal B} (-E)\rangle$, where $\exp(-i k \pi)=\exp( i k \pi)$. Consequently, the states $|{\cal B} (E)\rangle$ and $|{\cal B} (-E)\rangle$ act as degenerate states under the unitary action $U$. To construct a dark state we consider a linear combination of these two states, such that
\begin{equation}
	|D\rangle = c_1 |{\cal B}(E)\rangle+ c_2 |{\cal B}(-E)\rangle  = \mid \downarrow\rangle_0 \otimes |\beta^\downarrow\rangle.
	\label{eq:Fdark}
\end{equation}
Using Eq. (\ref{eq:Bright}), we have $c_1 = (2E-w_0) \sqrt{\langle \beta^\downarrow  |\beta^\downarrow\rangle+(E+w_0/2)^2 \langle \beta^\uparrow  |\beta^\uparrow\rangle}/(4E)$ and $c_2 =(2E+w_0)\sqrt{\langle \beta^\downarrow  |\beta^\downarrow\rangle+(w_0/2 -E)^2 \langle \beta^\uparrow  |\beta^\uparrow\rangle}/(4E)$. For the constructed state $|D\rangle$, when the sampling time satisfies the condition in Eq. (\ref{eq:darkc}), we obtain
\begin{equation}
 S |D\rangle =  e^{ -i k \pi} |D\rangle = e^{ i k \pi} |D\rangle.
 \label{eqS9}
\end{equation}
Here, $|D\rangle$ is an eigenstate of the survival operator with an eigenvalue $|\xi| = |\exp(-i k \pi)|=1$. Consequently, $|D\rangle$ also qualifies as a dark state. When $\tau \neq k\pi/E$, we find that $ S |D\rangle = D_{\downarrow} [ c_1 \exp(-i E\tau) |{\cal B}(E)\rangle+ c_2 \exp(i E\tau) |{\cal B}(-E)\rangle]$. Therefore, the state $|D\rangle$ is not an eigenstate of $S$ under these conditions. In other words, $|D\rangle$ only emerges for specific sampling times given in Eq. (\ref{eq:darkc}). Using Eq. (17), we can infer that the recurrence time will exhibit a sharp jump when these dark states appear.

The above discussion is generic for any central spin models. As an example, we confirm Eqs. (\ref{eq:darkc}-\ref{eqS9}) with the four-spin case. As shown in Fig. \ref{figsub2}, we have ten distinct energy levels, five with $E < 0$ and five with $E>0$. We label energy levels as $E_0 < E_1 < E_2 < E_3 < 0$. The states corresponding to $E_4$ are typical dark states as we discussed in the main text, i.e., independent of sampling time $\tau$. As there are typically three dark states, we have $\overline{n} = 13/8$, which is confirmed with numerical simulations as shown in Fig. \ref{figsub2}(B) for most of $\tau$s except the resonances. For the resonances, when $\tau$ increases, energy levels begin to satisfy the condition in Eq. (\ref{eq:darkc}). The energy level $E_0$ possesses the largest absolute value, and when $\tau = \pi/|E_0|$, a new dark state is constructed using Eq. (\ref{eq:Fdark}). Consequently, instead of three dark states we now have four dark states in total and $\overline{n} = 2-4/8=3/2$, which is in agreement with the results in Fig. \ref{figsub2}(B). When $\tau = \pi/|E_1|$, two additional dark states are constructed with Eq. (\ref{eq:Fdark}) due to the two-fold degeneracy of the energy level $E_1$, leading to $\overline{n} = 11/8$. In a similar fashion, all the jumps in the recurrence time can be explained using the rise of the dark states. A special case occurs when $\tau = 4\pi/|E_1|= 3\pi/|E_2|$, where energy levels $E_1$ and $E_2$ simultaneously satisfy Eq. (\ref{eq:darkc}). In this case, three additional dark states are constructed, resulting in $\overline{n} =5/4$.

%

\section{Counting the number of dark states}
\label{ap3}

In this section, we study the number of dark states for the central spin model Eq. (\ref{eq:hce}). We derive Eq. (20) in the main text. 


{\em No excitations.} We start with the ground state, denoted $| \psi \rangle_0=\mid\downarrow,\downarrow, \cdots\rangle$. Here, all the spins are in the ``down" state, i.e., the vacuum state. The subscript $0$ in $|\psi\rangle_0$ denotes zero spin excitations. It is easy to check that this is an eigenstate of $H$. This is because $\sum_{i=1} ^{N-1} g_i S_0 ^{+} S_i ^{-} |\psi\rangle_0 = 0$ and $S_0 ^{-}|\psi\rangle_0=0$. We have a unique disentangled state when the number of spin excitations in the sector is zero. By sector, we mean that here we study only the cases where the central spin is in the ``down" state. The energy of the state $|\psi\rangle_0$ is $- \omega_0/2$. Clearly, this state, being an eigenstate of $H$, is a dark state, as any click will detect the central spin in the state ``down".

{\em One excitation states}. We now find disentangled dark states with a single excitation of the bath states, denoted $|\psi\rangle_1$, where the subscript is for the number of excitations. For example, consider the case of four bath spins. We will find an eigenstate of $H$ such that

\begin{equation}
|\psi\rangle_1 = \alpha_1 \mid \downarrow, \uparrow,\downarrow,\downarrow,\downarrow\rangle+
\alpha_2 \mid \downarrow, \downarrow, \uparrow,\downarrow,\downarrow\rangle+ 
 \alpha_3 \mid \downarrow, \downarrow, \downarrow,\uparrow,\downarrow\rangle+
\alpha_4 \mid \downarrow, \downarrow, \downarrow,\downarrow,\uparrow\rangle.
\label{eqBethe01}
\end{equation}
Again, since this is a stationary state and the central spin is in state ``down", this state will remain forever dark, and the string of measurements will be down, down, down, till infinity. The condition that this holds is, as before, $\sum_{i=1} ^{N-1} g_i S_0 ^{+} S_i ^{-} |\psi\rangle_1 =0 $. Define two vectors, the first composed of the unknowns $\vec{\alpha} = (\alpha_1, \cdots, \alpha_{N-1})$ and the second from the prescribed couplings $\vec{g}= (g_1, \cdots, g_{N-1})$. Then we clearly need to solve $\vec{\alpha} \cdot \vec{g} = 0$. Since $\vec{g}$ is a vector of length $N-1$ given by the coupling constants, we know from basic geometry that we have $N-2$ vectors orthogonal to $\vec{g}$, which are also mutually orthogonal. As usual, they can also be normalized, but this is of little concern here. We reject the trivial solution $\vec{\alpha}=0$ since it does not represent a state. The important conclusion for us is that we have $N-2$ dark states with a single bath excitation. We remark that the energy of this state is again, $-\omega_0/2$, and the same holds for all the dark states considered below. The central spin and bath spins are not entangled because, in general, the dark state is $|D\rangle = \mid \downarrow\rangle \otimes |\psi\rangle_i$, which is separable.

{\em Two spin excitations.} We now consider dark states $|\psi\rangle_2$ with two spin excitations, for example, for four bath spins, we have
\begin{equation}
|\psi\rangle_2 = \alpha_{12} \mid \downarrow, \uparrow,\uparrow,\downarrow,\downarrow\rangle+
\alpha_{13} \mid \downarrow, \uparrow, \downarrow,\uparrow,\downarrow\rangle+
\alpha_{14} \mid \downarrow, \uparrow, \downarrow,\downarrow,\uparrow\rangle+
\alpha_{23} \mid \downarrow, \downarrow, \uparrow,\uparrow,\downarrow\rangle+
\alpha_{24} \mid \downarrow, \downarrow, \uparrow,\downarrow,\uparrow\rangle+
\alpha_{34} \mid \downarrow, \downarrow, \downarrow,\uparrow,\uparrow\rangle.
\label{eqBethe02}
\end{equation}
Like before, the central spin is in a state ``down", so this is a dark state. The matrix element, $\alpha_{ij}$, indicates that the $i$-th and the $j$-th bath spins are excited, so these indices run from $1$ to $N-1$. Clearly, the diagonal elements $\alpha_{ii}$ are zero, and $\alpha_{ij}= \alpha_{ji}$. In this example, with $4$ spins in the bath, we have six unknowns, $\alpha_{12},\alpha_{13},\alpha_{14},\alpha_{23},\alpha_{24}$ and $\alpha_{34}$. As before these states are dark eigenstates of $H$ if $\sum_{i=1} ^{N-1} g_i S_0 ^{+} S_i ^{-} | \psi\rangle_2 = 0$. This means that for four bath spins, we must solve
\begin{equation}
\left(
\begin{array}{c c c c}
0 & \alpha_{12} & \alpha_{13} & \alpha_{14} \\
\alpha_{12} & 0 &  \alpha_{23} & \alpha_{24} \\
\alpha_{13} &  \alpha_{23} & 0 & \alpha_{34} \\
\alpha_{14} &  \alpha_{24} & \alpha_{34} & 0 
\end{array}
\right) 
\left(
\begin{array}{c}
g_1 \\
g_2 \\ 
g_3 \\
g_4 
\end{array}
\right)=0.
\label{EqBethe03}
\end{equation}
Clearly, in this example, we have $4$ equations and $6$ unknowns, and hence we find two independent solutions. Technically, we may choose two coefficients as seeds, for example, $\alpha_{34},\alpha_{24}$. If these are set to zero, a short calculation shows that the other $\alpha_{ij}$ must be zero as well; however, this does not describe a legitimate state. Hence, we have two solutions that can be easily found by assigning say $\alpha_{34}=1, \alpha_{24}=0$ and vice versa.  Other pairs of seeds can be assigned, and then, with simple linear algebra, namely using the Gram-Schmidt orthogonalization, we find the constants $\alpha_{ij}$ and the orthogonal states. Again, this is not our aim here; our focus is that for $N=5$ we have two dark states. More generally, the number of dark states is the number of matrix elements above (or below) the diagonal of an $N -1 \times N -1$ matrix, minus the number of constraints, namely the dimension of the vector $\vec{g}$, which is $N-1$. Hence
\begin{gather}
	\mbox{number of dark states with a pair of excitations =}  \nonumber 
	\left(
 \begin{array}{c}
N-1 \\
2 
\end{array}
\right) - 
\left(
\begin{array}{c}
N-1 \\
1
\end{array}
\right) = { (N-4)(N-1) \over 2}.
\label{eqBethe4}
\end{gather}
Here we assumed that $N > 4$; otherwise, the number of dark states with two excitations is zero.

Eq. (\ref{eqBethe4}) has a simple combinatorial meaning. For two spin excitations, among $N-1$  bath spins, we have $\left(\begin{array}{c} N-1 \\ 2 \end{array} \right)$ ways to arrange the spins. This is the number of $\alpha_{ij}$s one needs to determine. On the other hand the operation of $\sum_{i=1} ^{N-1} g_i S_0 ^{+} S_i ^{-}$ on the two spin excitation states $|\psi\rangle_2$ is clearly to take the central spin denoted with $0$ from the ``down" state to ``up", while annihilating one of the two excitations. This means that for the bath spins, we are left with one spin in the state ``up".  The number of such states is clearly $\left(\begin{array}{c} N-1 \\ 1 \end{array} \right)$. This is the reason why Eq. (\ref{eqBethe4}) holds. To see this, note that $\sum_{i=1} ^{N-1} g_i S_0 ^{+} S_i ^{-} | \psi\rangle_2=0$  gives a linear combination of states with a single spin bath excitation, however these are all orthogonal, so multiplying by kets composed of these $N-1$ single bath spin excitation vectors, will give $N-1$ equations to determine the unknowns $\alpha_{ij}$. This leaves 
$\left(
 \begin{array}{c}
N-1 \\
2 
\end{array}
\right) - 
\left(
\begin{array}{c}
N-1 \\
1
\end{array}
\right)$
unconstrained $\alpha_{ij}$s. Let us denote this number by $N_{2 D}$. We choose $N_{2 D}$ members of the set $\alpha_{ij}$. We assign to these unconstrained $\alpha_{ij}$s orthogonal vectors [like what we did for $(\alpha_{34},\alpha_{24})\rightarrow (1,0)$ and $(0,1)$], but now we have $N_{2D}$ such vectors instead of two]. From these seed vectors, we find in principle the other $\alpha_{ij}$. Hence, $N_{2D}$ is also the number of dark states for two bath spin excitations.

{\em Three excited spins}. 
The number of ways to arrange $2$ spins among $N-1$ is $\left( \begin{array}{c} N-1 \\ 2 \end{array} \right)$, this being the number of equations for the $\alpha_{ijk}$ to be solved. The number of unknowns (number of possible three-spin excitations) is $\left( \begin{array}{c} N-1 \\ 3 \end{array} \right)$. Hence 
\begin{gather}
	\mbox{number of dark states with three excitations =  } 
	\left( \begin{array}{c} N-1 \\ 3 \end{array} \right)
-
 \left( \begin{array}{c} N-1 \\ 2 \end{array} \right).
\label{eqBethe6}
\end{gather}
For $N=5$ we get a negative number, so there are no dark states in this case. A similar cutoff will be found in general for other values of $N$.

{\em $K$ excited spins}. In general, we see that for a bath with $K$ excited spins, hence clearly $K \le N$, we have 
\begin{gather}
\mbox{ number of dark states with $K$ excitations =}  
 \left( \begin{array}{c} N-1 \\ K \end{array} \right)
-
 \left( \begin{array}{c} N-1 \\ K -1  \end{array} \right).
\label{eqBethe7}
\end{gather}
Clearly, the number of dark states cannot be negative. Thus, the maximum $K$, beyond which we cannot find dark states,  is easily shown to be $K_{{\rm max}}= N/2$ if $N$ is even, or $K_{{\rm max}} = (N-1)/2$ if $N$ is odd. This result can also be derived using the Bethe Ansatz treatment \cite{PhysRevLett.91.246802,Villazon2020}.   We now find the total number of dark states, summing over the number of dark states for the different values of excited spins in the bath
\begin{gather}
	\underbrace{\left( \begin{array}{c} N-1 \\ 0 \end{array} \right)}_{0\ \mbox{excitations}}  +
\underbrace{\left( \begin{array}{c} N-1 \\ 1 \end{array} \right)-
\left( \begin{array}{c} N-1 \\ 0 \end{array} \right)
}_{1\  \mbox{excitations}} + 
\underbrace{\left( \begin{array}{c} N-1 \\ 2 \end{array} \right) -
\left( \begin{array}{c} N-1 \\ 1 \end{array} \right)}_{2 \ \mbox{excitations}}
 + \cdots
\underbrace{\left( \begin{array}{c} N-1 \\ K_{{\rm max}} \end{array} \right)-
\left( \begin{array}{c} N-1 \\ K_{{\rm max}} -1 \end{array} \right)}_{K_{{\rm max}} \ \mbox{excitations}}
\end{gather}
Hence we conclude
\begin{equation}
	 \mbox{total number of  dark states =} \ \ 
\frac{\Gamma (N)}{\Gamma \left( \lfloor\frac{N+2}{2}  \rfloor \right) \Gamma \left( \lfloor\frac{N+1}{2}  \rfloor \right)2}.
\label{eqNUdark}
\end{equation}
Together with Eq. (17), we get the mean recurrence time versus $N$ for the central spin model, as shown in Eq. (20) in the main text. Note that this equation is not valid at resonances as previously discussed.

\begin{figure}[!t]
    \centering
    \includegraphics[width=0.35\columnwidth]{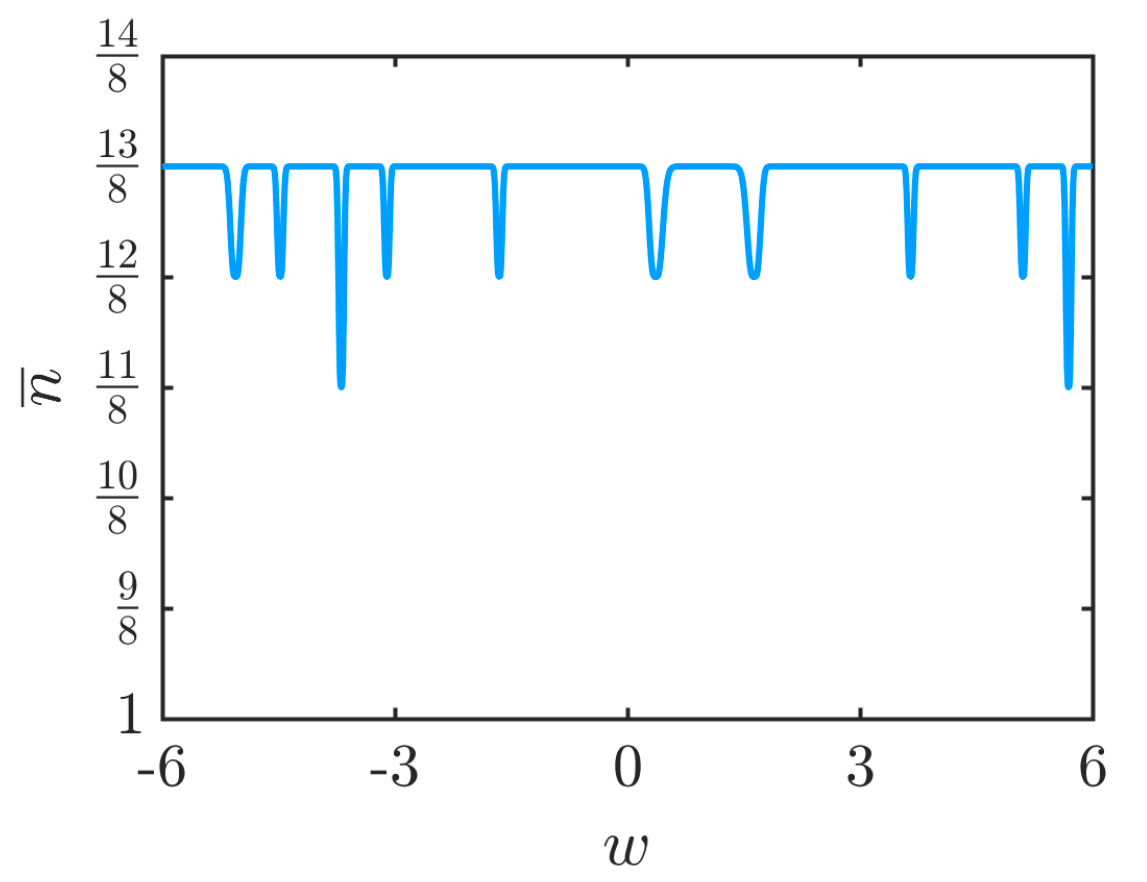}
    \caption{The ensemble mean $\overline{n}$ versus the magnitude of the magnetic field. Here we use the central spin model given in Eq. (\ref{eq:hce}) and $g_1 = 1.221$, $g_2 = 0.864$, and $g_3 = 1.46$. The sampling time $\tau=1$ is a constant.} 
    \label{figsub3}
\end{figure}

We noted already that the energy of all the dark states is $-\omega_0/2$ if $\omega=0$. If $\omega\neq 0$ in Eq. (\ref{eq:hce}), the degeneracy of these states would be partially removed, though states with $K$ excitations will have the same energy. However, the eigenfunctions would remain unchanged, hence also when $\omega\neq 0$, the number of dark states is given in Eq. (\ref{eqNUdark}). This in turn implies that a magnetic field $\omega$ will not modify the fractional mean return time, even if the degeneracy is lifted, see Fig. \ref{figsub3}.

\section{XXX central spin model}
\label{ap4}

In the finite number of measurements section of the main text, we utilize the XXX central spin model. The Hamiltonian of the XXX central spin model is given by:

\begin{equation}
    H = w_0 S_0^z + w \sum_{i=1}^{N-1} S_i^z + \sum_{i=1}^{N-1} g_i (S_0^xS_i^x + S_0^yS_i^y + S_0^zS_i^z).
\end{equation}

Here, the central spin $S_0$ is subject to an external magnetic field $w_0$, and the bath spins are subject to a magnetic field of strength $w$. This model differs from the XX central spin model defined in the main text, as the coupling between the central spin and bath spins occurs in the $X$, $Y$, and $Z$ directions.

The model is classified into isotropic or anisotropic cases, depending on whether the coupling constant $g_i$ is uniform (for instance, $g_i = \text{constant}$, with all $g_i$ equal to one) or varied (e.g., $g_i = 1$, $g_2 = 1.5$, etc.). For the isotropic XXX central spin model, the ensemble mean recurrence time as a function of system size is the same as that in the XX central spin model discussed in the main text. However, it should be noted that for the XX central spin model, Eq.~(20) is valid regardless of whether $g_i$ values are identical or varied.

For the anisotropic XXX central spin model, the ensemble mean recurrence time as a function of system size is expressed as:

\begin{equation}
    \overline{n} = 2 - \frac{1}{2^{N-1}}.
    \label{eqSnew}
\end{equation}

This equation mirrors the scaling behavior observed in the Heisenberg spin chain. Furthermore, these results suggest that the typical behaviors we presented in the main text are representative and applicable to other models.

\section{Full knowledge measurement and partial knowledge measurement}

\begin{figure}[!htbp]
    \centering
    \includegraphics[width=\columnwidth]{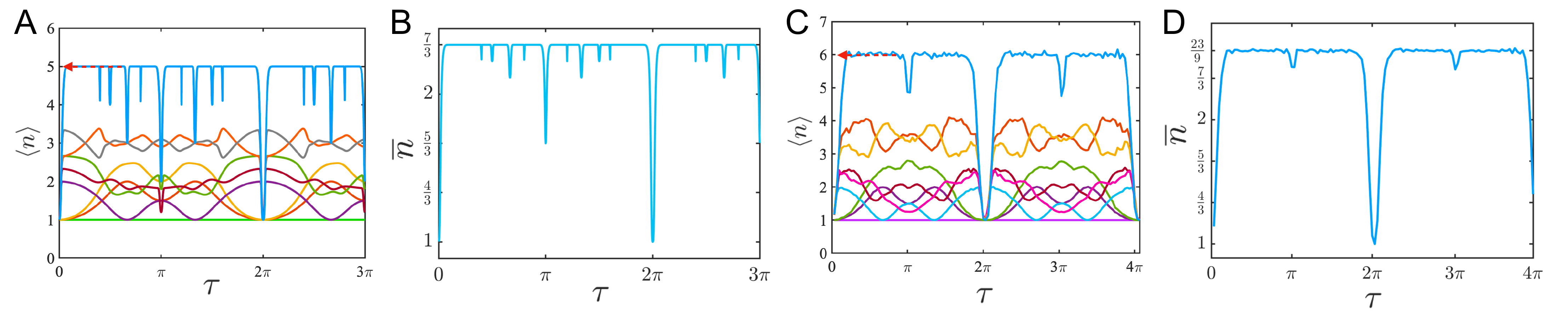}
    \caption{Mean recurrence time for a spin-1 chain under full and partial knowledge measurements for $N=3$. Here, the model used is the Heisenberg chain with open boundary conditions, like in Fig. 2, but now with spin-1 instead of spin-$\frac{1}{2}$. The measurements are performed on the first spin. (A,B) Full knowledge measurements: (A) the mean recurrence time for product initial states $|\psi_0\rangle$ is plotted as a function of the sampling time $\tau$. (B) The ensemble-averaged mean recurrence time for the spin-1 chain with its fractional quantization. (C,D) Partial knowledge measurements: (C) the mean recurrence time for product initial states $|\psi_0\rangle$ versus sampling time $\tau$ with recurrence times obtained via Monte Carlo simulations.  (D) The ensemble-averaged mean recurrence time shows a different fractional quantization. }    
    \label{fign3new}
\end{figure}

Fig.~\ref{fign3new} complements the main text analysis by presenting recurrence statistics for an open-boundary Heisenberg chain [Eq.~(1)] with spin-1 degrees of freedom, monitored on the first site at stroboscopic times. Two measurement protocols are considered. (i) \emph{Full knowledge}: a three-outcome projective measurement resolves $S_1^z\in\{+1,0,-1\}$ with projectors $D^{m}=\sum_{b}|m,b\rangle\langle m,b|$ with $m \in \{+1,0,-1\} $. (ii) \emph{Partial knowledge}: a two-outcome measurement distinguishes $|+1\rangle$ from $\{|0\rangle,|-1\rangle\}$ using $D^{+1}$ and $D_{0,-1}$.

Fig.~\ref{fign3new}(A,B) show the full-knowledge case, which exhibits fractional plateaus away from resonances. Fig.~\ref{fign3new}(C,D)  display the partial-knowledge case, where $\langle n \rangle $ is obtained from quantum-trajectory Monte Carlo and the corresponding $\overline{n}$ shows a distinct fractional quantization. For the specific state $|\psi_0\rangle=|+1,-1,-1\rangle$, integer-valued plateaus appear under both protocols, but with different values, reflecting the different effective Hilbert spaces probed by the measurements. Excluding resonant $\tau$, the ensemble plateaus take the representative values $7/3$ (full knowledge) and $23/9$ (partial knowledge), illustrating that the fraction depends on the information gained per measurement.

For the partial-knowledge protocol, define the survival step operator $S=D_{0,-1}U(\tau)$ with $U(\tau)=e^{-iH\tau}$. Using Eq.~(15) in the main text, the ensemble mean can be written in terms of the spectrum $\{\xi_j\}$ of $S$,
\begin{equation}
\bar n =
\frac{ \text{number of eigenvalues } \xi_j \text{ satisfying } |\xi_j|<1}{3^{N-1}}
\le 3,
\end{equation}
which yields the spin-1 upper bound in Eq.~(22) in the main text. The bound equals the local dimension $d=3$ and generalizes to spin-$X$ (local dimension $2X{+}1$) under the same partial-knowledge measurement, giving $\overline{n}\le 2X{+}1$. Overall, Fig.~\ref{fign3new} demonstrates that fractional quantization of the mean recurrence time persists in spin-1 chains and that its value is protocol dependent, while remaining bounded by the local on-site Hilbert-space dimension.

\end{document}